\documentstyle[prl,aps,12pt]{revtex}
\input epsf.sty

\newcommand{\eq}{\begin{equation}}
\newcommand{\en}{\end{equation}}
\newcommand{\BEF}{\begin{figure}}
\newcommand{\EF}{\end{figure}}
\newcommand{\bea}{\begin{eqnarray}}
\newcommand{\eea}{\end{eqnarray}}

\newcommand{\J}{\J_{\mu ab}}
\newcommand{\T}{\T^{\mu\nu}}

\newcommand{\unity}{1\kern-.65mm \mbox{\form l}}
\newfont{\form}{cmss10}
\let\varkappa\kappa

\begin{document}
\draft
\renewcommand{\baselinestretch}{1.5}
\pagestyle{empty}
\title{\Large\bf Chiral solitons from dimensional reduction of 
Chern-Simons gauged non-linear Schr\"odinger model of FQHE: 
classical and quantum aspects\footnote{This work is
  supported by the U.S. Department of Energy (D.O.E.) under
  cooperative  agreement \#DE-FC02-94ER40818, by NSF grant
  PHY-9315811, and by Istituto Nazionale di Fisica Nucleare (INFN,
  Frascati, Italy).}}
\bigskip
\bigskip
\author{L. Griguolo$^{(a)}$\footnote{e-mail:\tt griguolo@irene.mit.edu}  
and D. Seminara$^{(b)}$\footnote{e-mail:\tt seminara@binah.cc.brandeis.edu}}

\address{{\it $^{(a)}$ Center for Theoretical Physics
Laboratory for  Nuclear Science and Department of Physics\\
Massachusetts Institute of Technology Cambridge, Massachusetts 02139, U.S.A.}\\
{\it $^{(b)}$ Department of  Physics, Brandeis University
    Waltham, MA 02254, USA}}
\date{Received \today}
\maketitle
\bigskip
\bigskip
\begin{abstract}

The soliton structure of a gauge theory recently proposed to describe 
chiral excitations in the Fractional Quantum Hall Effect is investigated.
 A new type of non-linear
derivative Schr\"odinger equation emerges as an effective description of the
system that supports novel {\it chiral} solitons. We discuss the classical
properties of solutions with vanishing and  non-vanishing boundary
conditions (dark solitons) and we explain their relation
to integrable systems. The  quantum analysis is also addressed in the 
framework of a semiclassical approximation improved by Renormalization
Group arguments. 
\end{abstract}
\vfill
\begin{flushright}
MIT-CTP-2578\\
BRX-TH-401
\end{flushright}
\newpage
\pagestyle{plain}
\section{Introduction}
\baselineskip .8truecm

One of the most fascinating aspects of classical field theory is the 
existence of travelling localized solutions of the non-linear equations 
describing physical systems: these {\it solitons}  find 
application in a very broad class of physical problems. From the quantum 
field theory point of view, 
(being exact non-perturbative solutions of the classical theory)
they are believed to carry 
information about the non-perturbative structure of the quantum theory. 
Their particle-like properties have an intuitive and reasonable
interpretation as bound states of the elementary excitations of the 
corresponding quantum field theory. For specific two-dimensional models this 
idea was confirmed by explicit computations in the framework of the 
semiclassical approximation \cite{DHN}, of  quantum inverse scattering 
\cite{FADE} and of other systematic expansions \cite{brodolo} while, 
in more recent years, conformal field theory techniques have led to
exact results 
for the $S-$matrix \cite{ZAMO}. Solitons also play  
fundamental role in recent non-perturbative developments of QFT (for a
recent review, see \cite{DIVE}), since they are seen as the ``duals'' 
of the elementary excitations appearing in the original Lagrangian
\cite{COLE}.

Solitons made their appearance in a completely different context some
years ago, namely in low-energy phenomenological applications to physical 
systems confined to a plane. An interesting class of gauge theoretical models, 
describing non-relativistic matter coupled to a Chern-Simons gauge 
field, was introduced by Jackiw and Pi \cite{JAPI1} to obtain a simple 
realization 
of non-relativistic interacting anyons. For an appropriate choice of the 
self-interaction potential, in the static case, one can reduce the 
Euler-Lagrange equations to the completely integrable Toda equation 
(in the non-Abelian situation) or Liouville (in the Abelian one) \cite{JAPI2}, 
with well-known soliton solutions. 
The dynamics of the system corresponds, in this 
case, to a dimensional reduction in time: recently \cite{RABE,NOI} there
has been instead considered a reduction in one spatial dimension. 
The physical reason lies in the hope that 
some characteristic of the model, in particular its fractional 
statistics, can be maintained by the related one-dimensional excitations and 
that, by a suitable modification,  chiral behaviour can be induced. 
These two features are in fact relevant in the phenomenological description 
of the edge states in the Fractional Quantum Hall Effect \cite{WEN}. 
Unfortunately the theory obtained in this way, originally proposed 
in \cite{RABE}, partially fails to achieve its goal, because as was shown in 
\cite{NOI} no statistical transmutation arises for  non-relativistic matter. 
On the other hand, as first observed in \cite{NOI}, a novel and interesting 
soliton structure is present there, finding its origin in the gauge coupling 
and in the chiral modification. Our investigation is a direct 
follow-up of the work begun in \cite{NOI}, and it is particularly devoted to 
the classical and quantum analysis of the solitons appearing in the theory. 
Actually, after realizing that the symmetry property of the original model 
is shared by a larger class of $1+1$ dimensional theories, we have chosen 
to study a slight generalization of the model of \cite{RABE,NOI},
that fits equally well into the dimensional 
reduction procedure. Surprisingly, as we shall see, quantum effects 
naturally lead one to include the more 
general interactions that we have introduced. 
As in the analogous $2+1$-dimensional 
family, integrability appears only for a suitable choice of the initial 
parameters (as observed in \cite{COREA}), relating our system to 
the Derivative Non-Linear Schr\"odinger equation (DNLS) \cite{DNLS}. 
Nevertheless soliton solutions exist for any value of the coupling constants.

We start in Sect. 2 by describing the dimensional reduction of the Jackiw-Pi 
model and its modification, based essentially on the introduction of a chiral 
boson. 
The gauge action turns out to be of $B-F$ type and we show that the system is 
equivalent to a general family of non-linear Schr\"odinger equations that do 
not possess Galilean invariance. In absence of a self-interaction potential 
for the matter, the model has non-relativistic scale invariance, 
an important ingredient in understanding some features of the theory: a 
suitable potential respecting the scale invariance can also be added.
We discuss the conserved charges and we derive, from 
symmetry considerations, some general properties of localized classical 
solution. In Sect. 3 we compute the solitons, first without and then
including a potential, and we discuss their properties: in particular
they appear 
to be chiral, existing only for a fixed sign of the total momentum, and 
presenting an interesting particle-like behaviour, inherited from the scale 
invariance. We also consider finite energy solutions with non-vanishing 
boundary condition at  (spatial) infinity (dark solitons), that possess 
opposite chirality with respect to the previous ones. 
We show that conserved charges must be carefully defined  when
non-trivial boundary conditions  are present and we derive the relevant energy-momentum 
dispersion relation.
In Sect.4, we discuss the quantization of the theory, 
paying particular attention to the relation between the solitons 
and the quantum bound states. Since the system is non-integrable for a generic 
choice of the coupling constants, it is 
difficult to obtain exact results, and we have explicitly solved the
Schr\"odinger equation only for the two-body problem. Nevertheless we
have obtained a 
perturbative solution, consistent with the Bohr-Sommerfield quantization of the 
solitons (in the weak-coupling limit) and the exact expression of the 
trace anomaly of the theory, showing that classical scale invariance is 
 destroyed, except  at the fixed point of the 
renormalization group, represented by DNLS. 
Combining this result with the semiclassical quantization of the soliton, 
we conjecture the general form of the energy for the $n$-body bound-state.
Finally , in Sect. 5 we present our conclusion and indicate future directions.

\bigskip
\section{Non-linear derivative Schr\"odinger equations from a gauge theory 
and their classical symmetries}
A nonrelativistic gauge field theory that leads to planar anyons
\cite{JAPI1} is the
nonlinear Schr\"odinger equation, gauged by a Chern-Simons field and governed 
by the Lagrange density
\begin{equation}
{\cal L}_{(2+1)} = \frac1{4\bar\varkappa}\epsilon^{\alpha\beta\gamma} 
\hat A_\alpha \hat F_{\beta\gamma}
 + i\hbar \Psi^* (\partial_t + i\hat A_0)\Psi - 
\frac{\hbar^2}{2m} \sum_{i=1}^2 \left| (\partial_i + i\hat A_i)\Psi\right|^2
- V(\Psi^{*}\Psi) \label{eq:1} \\
\end{equation}
Here $\Psi$ is the Schr\"odinger quantum field, giving rise to charged bosonic 
particles after second quantization. $\hat A_\mu$ possesses no propagating 
degrees of freedom; it 
can be eliminated, leaving a statistical Aharonov-Bohm interaction  between the
particles. $V(\Psi^{*}\Psi)$ describes possible nonlinear self-interactions, 
and can be a general polynomial in the density $\Psi^{*}\Psi$. We notice that 
the above Lagrangian is invariant under Galilean transformations, due to the 
topological nature of the Chern-Simons action. 
When analyzing the lineal problem \cite{NOI}, it is natural to
consider a dimensional 
reduction of (\ref{eq:1}), by suppressing dependence on the second spatial 
coordinate, renaming $\hat A_2$ as $(mc/\hbar^2) B$ and redefining the gauge 
field as $\hat A_x=A_x$ and $\hat A_0=A_0+m c^2/2\hbar B^2$. 
In this way one is led to a $B$--$F$ gauge theory coupled to a
non-relativistic bosonic field in $1+1$ dimensions:  
\begin{equation} 
 {\cal L}_{(1+1)} = \frac1{2\varkappa} B \epsilon^{\mu\nu} F_{\mu\nu}
+ i\hbar \Psi^*(\partial_t + iA_0)\Psi - 
\frac{\hbar^2}{2m} \left| (\partial_x + iA_x)\Psi\right|^2
- V(\Psi^{*}\Psi) ,\label{eq:2}
\end{equation}
where $\varkappa \equiv (\hbar^2/mc)\bar\varkappa$ is 
dimensionless and we have neglected $\partial_x (B^3/3\hbar \varkappa)$ 
since it is a total spatial derivative. 
Eliminating the $B$ and $A_\mu$ fields decouples them completely, 
in the sense that the phase of $\Psi$ may be adjusted so that the 
interactions of the $\Psi$ field are solely determined by  $V$, and 
particle statistics remain unaffected \cite{observation}. In fact
 the equation of motion
obtained by varying $B$, $F_{\mu\nu}=0$, implies that the gauge field
$A_\mu$ is a pure gauge and it can be reabsorbed in a phase redefinition
of $\Psi$. 
%However let us notice that this elimination cannot be completely
%safe if the spacial coordinate spans a circle. In that case, upon eliminating
%the gauge field, the  potential $V(\Phi^*\Phi)$ is still the only interaction,
%but instead of a periodic $\Psi$, we will have
%\begin{equation}
%\label{putto}
%\Psi(x+2\pi R,t)=\exp(2\pi i a R)\Psi(x,t).
%\end{equation}
%Here $\displaystyle{a=\int_0^{2 \pi R} d x A_x}$ and $R$ the radius of the 
%circle\footnote{$a$ is independent of time because $F_{\mu\nu}=0$}. The 
%boundary condition (\ref{putto}) is not a simpthom of non trivial statistics.
%It is simply the price  paid  in order to gauge away the flat connection
%associated with $S^1$.
%Choosing the gauge $A_{x}=0$, for 
%example, the equation of motions leads to $A_{0}=-{mc^2\over 2\hbar^3}B^2$ 
%and no extra-interaction appears in the Schr\"odinger equation apart it from 
%the non-linear term derived from $V(\Psi^{*}\Psi)$.
In order to obtain a non-trivial theory even in absence of $V$ it is 
quite natural to introduce a kinetic term for $B$, which, for example, 
could be taken to be in the Klein-Gordon form.  In the following, however,
we prefer a simpler expression that describes ``chiral'' Bose fields,
propagating only in one direction. The reason why we commit ourselves
to this specific choice is twofold: we hope both to induce a 
statistical transmutation and that the chiral dynamics is inherited by
$\Psi$. Having a field theoretical model that was both chiral and with
anomalous statistics would be of great relevance in the description of
the edge states \cite{WEN}. As
we shall see, while this model achieves the first goal (chirality), it
fails the other one. On the other hand its structure is so rich that a 
detailed  study of its properties has intrinsic interest.
  
The Lagrangian density for the chiral boson is proportional to
 $\pm \dot B B' - v B' B'$ \cite{ref:3}. (Dot/prime indicate
differentiation with respect to time/space.) Here $v$ is a velocity and the 
ensuing equations of motion for this kinetic term 
(without further interaction) are solved by $B=B(x\pm vt)$ 
(with suitable boundary conditions at spatial infinity), describing 
propagation with velocity $\mp v$.  Note that $\dot B B'$ is not invariant 
against a Galileo transformation, which is 
a symmetry of ${\cal L}_{(1+1)}$ and of $B' B'$: performing a Galileo boost 
on $\dot B B'$ with velocity $\tilde v$ gives rise to $\tilde v B' B'$, 
effectively boosting the $v$~parameter by
$\tilde v$. Consequently one can drop the $ v B'B'$ contribution to the 
kinetic $B$ Lagrangian, thereby selecting to work in a global 
 ``rest frame". Boosting a solution in this rest frame then produces a 
solution to the theory with a $B'B'$ term. 
In view of the above, we choose the $B$-kinetic Lagrange density to be
\begin{equation}
{\cal L}_B = {\lambda\over 2\varkappa^2\hbar} \dot B B' \label{eq:3}
\end{equation}
and the total Lagrange density is ${\cal L} = {\cal L}_B +
{\cal L}_{(1+1)}.$ The ensuing equations of motion are
\begin{mathletters}
\begin{eqnarray}
\label{eq:5a}
&&i\hbar(\partial_{t}+iA_{0})\Psi+{\hbar^2\over 2m}
(\partial_{x}+iA_{1})^2\Psi-V'\Psi=0,\\
&& F_{01}-\frac{\lambda}{\varkappa\hbar} \dot B'=0,
\label{eq:5b}\\
&&B'-\hbar \varkappa \Psi^{*}\Psi=0,
\label{eq:5c}\\
&&\dot B+\hbar \varkappa \hat J=0,
\label{eq:5d}
\end{eqnarray}
\end{mathletters}%
with $\hat J={\hbar\over 2im}\Bigl(\Psi^{*}(\partial_{x}+iA_{1})\Psi
-\Psi(\partial_{x}-iA_{1})\Psi^{*}\Bigr)$.  The integrability
condition for the last two equation leads to the usual  continuity 
equation 
\begin{equation}
\partial_{t}(\Psi^{*}\Psi)+\partial_{x}\hat J=0. 
\label{eq:7}
\end{equation}
Now in terms of  field
$\displaystyle{
\hat\phi=\exp\left (i\int^{x}_{x^0}~dy~ A_{1}(y,t)
+i\int^{t}_{t^0}~dt^\prime~ A_{1}(x^0,t^\prime)
-i\frac{\lambda}{\varkappa\hbar}B(x^0,t)\right)\Psi}$
eqs (\ref{eq:5a}) and (\ref{eq:5b}) become respectively
\begin{eqnarray}
&&i\hbar\partial_{t}\hat\phi+{\hbar^2\over 2m}\partial_{x}^2\hat\phi-\hbar 
{\lambda}j\hat\phi-V'\hat\phi=0,
\label{eq:8}\\
&&F_{01}={\lambda}\partial_{t}(\hat\phi^{*}\hat\phi).
\label{eq:9}
\end{eqnarray}
The latter gives the electromagnetic field as a function of 
the density $\hat\phi^{*}\hat\phi$ while the former encodes all the dynamical
content of the system.  In the following we shall be mainly interested 
in the case in which the potential $V(\hat\phi^*\hat\phi)$ is absent: this
 corresponds 
to a Schr\"odinger equation with a {\it current\/} 
 ($j$) nonlinearity: 
\begin{equation}
j= {\hbar \over 2im} \left( \hat\phi^{*}\partial_x\hat\phi -
\hat\phi\partial_x\hat\phi^{*} \right). 
\nonumber
\end{equation}
This is to be contrasted with the familiar nonlinear Schr\"odinger equation, 
where the nonlinearity involves the {\it charge density\/}
($\hat\phi\hat\phi^*$):
\begin{equation}
 i \hbar\partial_t\hat\phi = -{\hbar^2 \over 2m} \partial_x^2\hat\phi -
\lambda(\hat\phi\hat\phi^*) \hat\phi .
\label{eq:10}
\end{equation}
Note that the gauge interaction has dynamically produced a non-trivial 
alternative to the usual non-linear Schr\"odinger equation, much studied 
both from physical and mathematical points of view. However eq. (\ref{eq:8}),
in contrast to eq. (\ref{eq:10}), does not possess a local Lagrangian 
formulation directly in terms of the field $\hat\phi$. A local Lagrangian
can be instead exhibited for the  equation
\begin{equation}
i\hbar \partial_t \phi = - {\hbar^2 \over 2m} 
\left(\partial_x +i\frac{\lambda}{2} \rho^2
\right )^2 \phi+\frac{\lambda\hbar}{2} J \phi+V'\phi,
\label{eq:11}
\end{equation}
governing  the gauge equivalent variable $\phi=\displaystyle{
\exp\left[ i\frac{\lambda}{2}\left( \!\int^x_{x^0}\!dy(\hat\phi^*\hat\phi)(y,t)-
\!\!\!\int^t_{t^0}\!dt^\prime j(x^0,t^\prime)\right)\right ]\hat\phi}$.
Consider, in fact, the action
\begin{equation}
S=\int dt\,dx{\cal L}=\int dt\,dx
\left [ \frac{i\hbar}{2}(\hat{\phi}^{*}\partial_{t}{\phi}-
{\phi}\partial_{t}{\phi}^{*})-
\frac{\hbar^2}{2m}\left|\left(\partial_x +i{\lambda\over 2}\rho^2
\right) \phi\right|^{2}-V(\rho^2)\right ];
\label{eq:14}
\end{equation}
the Euler-Lagrange equations that follow can be easily shown to 
reproduce eq. (\ref{eq:11}). In eqs. (\ref{eq:11}) and (\ref{eq:14})
$\rho^2$ represents the density $\phi^*\phi$, while 
\begin{equation}
J={\hbar \over 2im} 
\left( \phi^{*} \left(\partial_x +i \frac{\lambda}{2}\rho^2\right) \phi -
\hat\phi \left(\partial_{x} - i \frac{\lambda}{2} \rho^2\right) \phi^{*}
\right)
\end{equation}
is the corresponding current.

The invariance of the action $S$ under 
space/time translations reflects itself into the presence of a divergenceless
energy momentum tensor $T^\lambda_{~\mu}$:
\begin{mathletters}
\begin{eqnarray}
&&T^{0}_{~0}={\cal H}=\Biggl[\frac{\hbar^2}{2m}\left|
D_x\phi \right|^2 +V(\rho^2)\Biggr]\ \ \ \
T^{x}_{~0}=-\frac{\hbar^2}{2m}
\left [ D_x \phi
\partial_t\phi^*
+(D_x \phi)^* \partial_t\phi
\right],\\
&&\nonumber\\
&&T^0_{~x}={\cal P}=m J-\hbar {\lambda\over 2} \rho^4\ \ \ \ \ \ \ \ \ \ \ \ 
\ \ \ \ \ T^{x}_{~x}= {\hbar^2\over m}\left| D_x  \phi \right|^2
-{\hbar^2\over 4m}\partial^{2}_{x}\rho^2
+V^{'}\rho^2-V,
\end{eqnarray}
\end{mathletters}
where $D_x$ stands for the ``covariant'' derivative $\partial_x
+i\lambda/2 \rho^2$. For solutions obeying suitable boundary conditions,
we can therefore write a conserved momentum $\displaystyle{P=\int dx~
{\cal P}}$ and energy $\displaystyle{E=\int dx~ {\cal H}}$. 
The unconventional form of the momentum density signals  lack of
Galilean invariance. In fact,  in contrast to the Galilean invariant case,
${\cal P}$ is not proportional to the $U(1)$ current  $J$. 
Indeed computing the time derivative of the Galileo generator
$\displaystyle{G=tP-m\int dx x\rho^2}$, we find
\begin{equation}
\label{urgo1}
\frac{d G}{d t}=\int dx ({\cal P}-m J)=-\hbar{\lambda\over 2} \int dx \rho^4,
\end{equation}  
namely $G$, depending on the sign of the coupling constant, always
increases or decreases in time. On the other hand, for a particular
choice of the potential $V$, the theory becomes scale invariant. In
fact the action (\ref{eq:14}) is unchanged under a dilation,
$t\rightarrow a^2 t,~x\rightarrow a x$, and ${\phi}(x,t)\rightarrow
a^{1\over 2}{\phi}(a^2 t, a x )$, iff $V({\rho^2/a}) a^3=V(\rho^2)$:
$\displaystyle{V(\rho^2)\propto\rho^6}$.
%[Here
%$\xi$ is a free dimensionless parameter.]
The generator $D$ of the scale symmetry takes the form
\begin{equation}
D=\int dx {\cal D}=tH-{1\over 2}\int dx x \cal{P},
\label{eq:dilaton}
\end{equation}
where the density ${\cal D}=t T^{0}_{~0}-{1\over 2} x T^{0}_{~x}$ obeys
the continuity equation
\begin{equation}
\partial_{t}{\cal D}+\partial_{x}\left (
t~T^{x}_{~0}-{1\over 2}x~T^{x}_{~x}-{\hbar^2\over 8m}
\partial_{x}\rho^2\right )=0.
\label{eq:xcontD}
\end{equation}
We can remove the last term in eq. (\ref{eq:xcontD})
proportional to the derivative of
$\rho^2$ by adding a superpotential to the  energy-momentum tensor.
In fact if we define an improved $\hat T^\lambda_{~\mu}$
\begin{equation}
\hat{T}^{0}_{~0}=T^{0}_{~0}-\displaystyle{{\hbar^2\over 8m}}
\partial^{2}_{x}\rho^2,\ \ \ \ \hat{T}^{x}_{~0}=T^{x}_{0}-
\displaystyle{{\hbar^2\over 8m}}\partial^{2}_{x}J,\ \ \ \
\hat{T}^{0}_{~x}=T^{0}_{~x},\ \ \ \
\hat{T}^{x}_{~x}=T^{x}_{~x},
\label{eq:improved}
\end{equation}
we obtain
\begin{equation}
\partial_t {\cal D}+\partial_x \left (t 
\hat T^x_{~0}-\frac{1}{2} x \hat T^{x}_{~x}\right )=0.
\end{equation}
The new energy  momentum tensor  satisfies $
2\hat{T}^{0}_{~0}=\hat T^{x}_{~x},$ 
which is the non relativistic analog of the relativistic traceless
condition for scale symmetry.  In a Galilean invariant theory  this condition
would also ensure conformal invariance; here, instead, the
absence of Galilean symmetry requires the absence of conformal symmetry.
Notice that the classical symmetries of our system form  a three
dimensional Lie sub-algebra of the conformal group (non-relativistic
Weyl group in $(1+1)-$dimensions) with Poisson brackets
\begin{equation}
\{H,P\}=0, \ \ \ \  \{D,P\}= P, \ \ \ \   \{D,H\}= 2 H.
\end{equation}
 
In the following we shall be interested in  classical ``localized''
solutions of the scale invariant Hamiltonian
\begin{equation}
\label{models}
H=\frac{\hbar^2}{2m}\int dx\left[\left|\left(\partial_x +i{\lambda\over 2}
\rho^2\right) \phi\right|^{2}-\frac{\lambda^2\xi}{4}
(\phi\phi^*)^3\right ] ,
\end{equation}
namely  normalizable field configurations ($\displaystyle{N=\int dx
\rho^2<\infty}$) with finite energy and momentum. Here $\xi$ is a
dimensionless parameter governing the strength of the cubic potential.
In the absence of $V$ ({\it i.e.} $\xi=0$) we recover the model of ref.
\cite{RABE,NOI} that will be our main concern and whose solutions
we shall  present in detail. Some results about $\xi\ne 0$ will be also
discussed as well because of their relevance in the quantum theory.

\indent
Unlike the conventional nonlinear Schr\"odinger equation, the family of
scale invariant equations we have described, and in particular the one without
cubic potential, does not appear to be completely integrable and thus 
analytic expressions for multisoliton solutions are not available. It has been 
remarked, however, that for $\xi=1$ the 
resulting Schr\"odinger equation becomes an integrable nonlinear, derivative 
Schr\"odinger equation with nonlinearity $i{\hbar^2\lambda^2\over 2 m}\rho
\partial_{x}\hat{\phi}$ \cite{COREA}.

\section{Classical Solutions}

When it comes to the problem of solving non-linear differential equations,
very few general tools are available. One possibility is to look for 
solutions that possess particular symmetries or whose specific functional 
dependence simplifies the structure of the original equation.  
In our case a simple Ansatz, which allows us  to integrate
eq. (\ref{eq:11}), is to assume that the density is a function only of $x-vt$, 
i.e. $\rho(x,t)\equiv \rho(x-v t).$
Drawing from the experience of integrable models, we can expect that this
choice will allow us to explore the presence of one-soliton solutions or 
more generally of traveling waves.  Substituting the ansatz  into the 
continuity equation, yields
\begin{equation}
\partial_x ( - v \rho^2(x-v t)+J(x,t))=0,
\end{equation}
%which can be promptly integrated
and hence 
\begin{equation}
\label{current1}
%J(x,t)-v \rho^2(x-v t)=J_\infty (t)\ \ {\rm or \ equivalently}\ \ \ 
J(x,t)= v\rho^2(x-vt) +J_\infty (t).
\end{equation}
Here $J_\infty(t)$ is an arbitrary function of time, whose physical
meaning will become transparent later. [ It is already clear
that such a function is related to the boundary conditions chosen at
infinity. ] In the following we shall be concerned  with solutions that 
approach  the vacuum at spatial infinity. Since ${\cal H}$ 
is positive definite in the absence of the potential $V(\rho^2)$, the
vacuum  solutions is first constrained by the requirement
\begin{equation}
H=0 \ \ \ \ \Longrightarrow \ \ \ \  
\Pi\equiv \left (\partial_x + i \frac{\lambda}{2}\rho^2\right) \phi=0,
\end{equation}
which is solved by taking $\phi$ of the form
\begin{equation}
\label{vacuum}
\phi(x,t)= \rho_0(t) \exp\left (-i \omega(t)- i \frac{\lambda}{2}
\rho_0^2(t) x \right ).
\end{equation}
Requiring (\ref{vacuum}) to satisfy the equation implies that  
$\rho_0(t)$ and $\omega(t)$  must be  constants. Thus
\begin{equation}
\label{vacuum1}
\hat \phi(x,t)_{vacuum}
=\rho_0\exp\left [-i \left (
\frac{\lambda}{2}\rho^2_0 x +\theta_0\right )\right ].
\end{equation}
Let us notice that the vacuum solution (\ref{vacuum1}) is characterized
by a vanishing current $J(x,t)$. Thus requiring that a solution approach
the vacuum  unambiguously fixes the value of $J_\infty(t)$ in eq. 
(\ref{current1}), namely 
\begin{equation}
\label{fincurrent}
J(x,t)=v (\rho^2(x-v t)-\rho^2_0).
\end{equation}
%[Let us remind to the reader that a  minimum of the energy satisfies
%automatically the equation of motion only if it is static.] 

\noindent
This brief analysis of the space of vacua suggests that we distinguish
two possible cases:  solutions that approach the trivial vacuum
$\phi_{vacuum}=0$ ($\rho_0=0$) at spatial infinity, and the solutions that, 
instead, go to the ground state that has constant density $\rho_0\ne 0$.
The former are characterized by well-defined energy, momentum and $U(1)$ 
charge. The latter have finite energy, but  infinite momentum
and $U(1)$ charge. This different behavior is easily understood in terms
of the properties of the vacuum. In fact, while in both cases the vacuum is
is invariant under time translation, the space translation and $U(1)$ 
transformation leave it invariant only if $\rho_0=0.$

\subsection{Solutions around the trivial vacuum $(\rho_0=0)$}

This class of solutions is strongly constrained by the symmetry of the
problem. Let us show how a wide number of their properties can be derived
without  explicitly solving  the equation of motion. To begin with, the
simple Ansatz $\rho\equiv\rho(x-v t)$, by means of eq. (\ref{fincurrent}),
implies that the momentum density ${\cal P}$ is a function of $x-v t$
as well.  Thus the dilation charge takes the form
\begin{equation}
D=t H -\frac{1}{2} \int^{\infty}_{-\infty} d x (x- v t) {\cal P}(x- v t)-
\frac{v}{2} t \int^{\infty}_{\infty} dx {\cal P}(x-v t)= t (H- \frac{v}{2} P) -\frac{1}{2}
D_0,
\end{equation}
where $D_0=\displaystyle{\int^{\infty}_{-\infty} dx x {\cal P}(x)}$. Since
$D$ is conserved and consequently time-independent we obtain
\begin{equation}
\label{HvP}
H=\frac{v}{2} P,
\end{equation}
namely the dispersion relation of a non-relativistic particle. Note that
the result holds independently of the value of $\xi$. In fact this reasoning
does not make use of the explicit form of the Hamiltonian, but only of its
symmetries.

To study further properties, we introduce the ``center of mass'' coordinate
\begin{equation}
x_{CM}(t)=\frac{\displaystyle{\int_{-\infty}^\infty dx~ x \rho^2(x,t)}}{
         \displaystyle{\int_{-\infty}^\infty dx~\rho^2(x,t)}}.
\end{equation}
This name is easily understood
if we think of $\rho^2$ as the mass density. Its velocity will be
\begin{equation}
v_{CM}=\dot x_{CM}(t)=\frac{\displaystyle{\int_{-\infty}^\infty dx~ J(x,t)}}{N},
\ \ \ \ \ {\rm with}\ \ N\equiv\int^{\infty}_{-\infty} dx ~\rho^2(x,t).
\end{equation}
Here we have used the continuity equation for the current to eliminate the
time derivative of the density. In this language the violation of 
Galilean  invariance (\ref{urgo1}) takes a suggestive form,
\begin{equation}
\label{purgo1}
\lambda (P- m N v_{CM}(t))=-\hbar\frac{\lambda^2}{2}\int^{\infty}_{-\infty}
dx ~\rho^4(x,t)\le 0.
\end{equation}
Being valid for all t, this implies
\begin{equation}
\lambda P\le m N \lambda \min_{t\in  {\rm R}}\{ v_{CM}(t)\}
\end{equation}
For $\xi<0$ ({\it i.e.} repulsive potential and thus energy positive
definite), one can show an analogous inequality for the energy. In
fact let us consider the following inequality
\begin{equation}
\int^\infty_{-\infty} dx \left | \phi + w \frac{\hbar}{2 m i}D_x \phi\right|^2
\ge 0,
\end{equation}
where $w$ is  an arbitrary parameter. In terms of the physical quantities
\begin{equation}
N + w N v_{CM} (t)+w^2 \frac{E_0}{2m} \ge 0.
\end{equation}
with $E_0=\displaystyle{\frac{\hbar^2}{2 m}
\int^\infty_{-\infty} dx \left | D_x \phi\right|^2}$. The fact that the
previous equation holds for all $w$  entails
\begin{equation}
E_0\ge \frac{m N v^2_{CM}(t)}{2},
\end{equation}
but, for $\xi<0$,the real energy $H$ is greater than $E_0$ and thus
\begin{equation}
\label{purgo3}
E\ge \frac{m N v^2_{CM}(t)}{2}.
\end{equation}
Let us apply these inequalities to our Ansatz. A simple
computation shows that $v_{cm}(t)=v$. Thus eq. (\ref{purgo1})
can be rewritten as
\begin{equation}
\lambda v\left( E-\frac{m N v^2}{2}\right)\le 0,
\end{equation}
where we used eq. (\ref{HvP}).
For $\xi<0$, because of (\ref{purgo3}), it implies $\lambda v<0$,
{\it i.e.} the soliton is ``chiral''. In other words, given the
sign of the coupling constant $\lambda$, the sign of the velocity
is determined. Recalling that
\begin{equation}
P=m N v -\hbar\frac{\lambda}{2}\int^\infty_{-\infty}\rho^4(x,t) dx,
\end{equation}
the chirality can be also written in the form $\lambda P<0$. In this
weaker form it is actually true for $\xi>0$ as well (see below).

%The assumption that $\hat\phi(x,t)\to 0$ at infinity is compatible
%with the equation  (\ref{current1}) only if we choose $J_\infty(t)=0$.
%[Actually, in order to prove the previous statement, we have implicitly
%assumed that our solution is regular, i.e. its modulus and phase are 
%differentiable bounded functions.] 
We return now to the problem of finding explicit solutions. Upon introducing
the parametrization
\begin{equation}
\label{parametrization}
\phi(x,t)=\rho(x-v t)\exp[i\theta(x,t)],
\end{equation}
eq. (\ref{fincurrent}) becomes  
\begin{equation}
\label{theta}
\partial_x\theta(x,t)=\frac{m v}{\hbar}-{\lambda\over 2}\rho^2(x-v t)\ \ \
\Longrightarrow\ \ \ \theta(x,t)=\frac{ m v}{\hbar}x+\theta_0(t)-
{\lambda\over 2}\int^{x- v t}_{-\infty}~dy~\rho^2(y),
\end{equation}
%where we have introduced the parameters $\alpha=\displaystyle{\hbar\over 2 m}$
%and $\beta=\displaystyle{-\hbar{\lambda\over m}}$  in order to simplify the 
%notation.
Substituting (\ref{theta})  into the equation of motion, we  obtain
\begin{equation}
\label{ccc}
\frac{2 m}{\hbar}\dot \theta_0(t) \rho(y)=\rho^{\prime\prime}(y)-
\frac{m^2 v^2}{\hbar^2}\rho(y)-{2 m v\over \hbar}\lambda \rho^{3}(y).
\end{equation}
Here $y\equiv x- v t$ and the prime denotes the derivative with respect to $y$. 
The consistency of eq. (\ref{ccc}) requires that $\dot\theta_0(t)$ is a 
constant $\omega_0$. Thus our equation can be rewritten as
% which in turn
%implies $\theta_0(t)=\omega_0 t+\theta_0$. Keeping in mind this last 
%result, our equation can be rewritten as follows
\begin{equation}
\rho^{\prime\prime}-
\frac{2m}{\hbar}\left (\omega_0+\frac{m v^2}{2\hbar }\right )\rho -
{2 m v\over \hbar}\lambda \rho^{3}=0,
\end{equation}
or equivalently 
\begin{equation}
\label{cop1}
(\rho^{\prime})^{2}-
\frac{m^2 v^2}{4 \hbar^2}\gamma\rho^2-
{m v\over \hbar}\lambda \rho^{4}=0,
\end{equation}
where $\displaystyle{\frac{m v^2}{8\hbar}\gamma=\left (\omega_0+
\frac{m v^2}{2\hbar }\right )}$.
The possible arbitrary constant integration in eq. (\ref{cop1}) is 
fixed by imposing that $\rho\to 0$ as $x\to\pm\infty$. The problem
of finding normalizable solutions is thus reduced to computing the 
zero energy orbits for a particle moving in a effective quartic 
potential. It is well-known that non-trivial ({\it i.e.} not identically 
constant) zero-energy orbits  exist if and only if
\begin{equation}
\label{condi}
\lambda v <0 \ \ \ \ \ \ \ {\rm and}\ \ \ \ \ \ \ \gamma >0.
\end{equation}
The first condition is particularly intriguing: given the sign of the
coupling constant, a solution can be found only for a given sign of $v$,
namely the system is {\it "chiral".} Integrating eq. (\ref{cop1}) in 
the allowed parameter region, we obtain
\begin{equation}
\rho(x-vt)=\frac{1}{2}\sqrt{~\left |\frac{m v}{\lambda\hbar}\right |\gamma}~
{\rm sech}\left[\frac{m |v|}{2\hbar}\sqrt{\gamma}
(x-x_0-v t)\right ]. 
\label{prisoli}
\end{equation}
The phase can be in turn computed with the help of eq. (\ref{theta})
\begin{equation}
\label{prisoli2}
\theta=\frac{m v}{\hbar} x+\omega_0 t+\theta_0-\frac{\sqrt{\gamma}}{4}
{\rm sign}(\lambda)
\tanh\left[\frac{m |v|}{2\hbar}\sqrt{\gamma}
(x-x_0-v t)\right ].
\end{equation}
The soliton's dynamical parameters, such as the $U(1)$ charge, the energy
or the momentum, can be now evaluted. Setting the  $U(1)$ charge
\begin{equation}
\label{Numero1}
N=\int^{+\infty}_{-\infty} \,dx\,\rho^2=\frac{\sqrt{\gamma}}{|\lambda|}.
\end{equation}
we find that the  energy and the momentum  of the  solitonic solution are
\begin{equation}
E={1\over 2}Mv^2,\ \ \ {\rm and} \ \ P=M v \ \ \ \ {\rm where} \ \  M=m N(1+{1\over 12}
\lambda^2 N^2).
\end{equation}
The soliton's characteristics are those of a non-relativistic particle of 
mass M, moving with velocity v and composed  of $N$ ``constituents''.
Notice that $N$ is a function of $v$ and $\omega_0$: for a given $N$ we can 
get solitons of arbitrary velocity simply by tuning the phase velocity.

As we have seen in the previous section many properties are shared by 
the whole class of scale invariant systems: it is therefore interesting to 
explore the one-soliton structure of our theory when a potential cubic in 
$\rho^2$ is present. We parametrized $V(\rho^2)$ as
\begin{equation}
V(\rho^2)=-{\hbar^2\lambda^2\xi\over 8 m}\rho^6,
\end{equation}
with $\xi\in(-\infty,+\infty)$. For $\xi=1$ we recover the integrable 
model described by the non-linear derivative Schrodinger equation, studied 
in \cite{COREA}. Taking into accounts vanishing boundary conditions for $\rho$ we 
simply have 
\begin{equation}
\label{cop2}
(\rho^{\prime})^2- \frac{m^2 v^2}{4\hbar^2} \gamma \rho^2-
\frac{m v}{\hbar}\lambda\rho^{4}+{\lambda^2\xi\over 4}\rho^6=0.
\end{equation}
After defining $\displaystyle{
Z={\hbar^2\lambda^2|\xi|\over m^2 v^2}\rho^2}$ and 
eq. (\ref{cop2}) becomes
\begin{eqnarray}
\label{cop3}
&&(Z^{\prime})^2={m^2 v^2\over\hbar^2 \xi}Z^2\left[(\gamma\xi+4)
-(Z-2~{\rm sign}(\lambda v))^2\right]\,\,\,\,\,\,\, \xi>0\nonumber\\
&&(Z^{\prime})^2={m^2 v^2\over \hbar^2 |\xi|}Z^2\left[(\gamma|\xi|-4)
+(Z+2~{\rm sign}(\lambda v))^2\right]\,\,\,\,\,\,\, \xi<0.
\end{eqnarray}
Let us first consider $\xi>0$: following the same arguments of $\xi=0$ case, we 
have normalizable solution for $\gamma>0$ that are
\begin{equation}
 \rho^2=|{m v\over\lambda\hbar}|{\gamma \over\displaystyle{ 
\sqrt{\gamma\xi+4}\cosh\left [{m v\over \hbar}\sqrt{\gamma}(x-x_0-vt)
\right ]-2{\rm sign}(\lambda v)}}.
\end{equation}
In this range the solitons are not chiral,  solutions exist for both the 
signs of the velocity: the condition on $\omega_0$ is exactly the same as  
when $\xi=0$. Moreover it is immediately realized that the limit 
$\xi\rightarrow 0$ reproduces the correct solution only for 
$\lambda v<0$, the other sign leading to a singular function. 
For $\xi<0$ the situation is rather different: a short analysis of the 
``fictitious'' potential yields the condition  
\begin{equation}
\label{condi1}
\lambda v <0 \ \ \ \ \ \ \ {\rm and}\ \ \ \ \ \ \
0<\gamma<{4\over|\xi|}.
\end{equation}
The functional form of $\rho^2$ is nevertheless similar,
\begin{equation}
 \rho^2=|{m v\over\lambda \hbar}|{\gamma \over\displaystyle{ 
\sqrt{4-\gamma\xi}\cosh\left[{m v\over\hbar}\sqrt{\gamma}(x-x_0-vt)\right ]+2}}.
\end{equation}
This  family of solutions is chiral and reproduces the solitons of the original 
equation. The different features of the two ranges and the particularity of the  
point $\xi=0$ are clearly displayed when the conserved quantities are computed:
\begin{mathletters}
\label{sol1}
\begin{eqnarray}
&&N={4\over |\lambda|\sqrt{\xi}}\arctan\sqrt{{\sqrt{4+\gamma\xi}+
2~{\rm sign}(\lambda v)\over \sqrt{4+\gamma\xi}-2 ~{\rm sign}(\lambda v)}},\\
&&E={m v^2\over 2|\lambda|}\left[{(\xi-1)\over \xi}{N\lambda}-
{2~{\rm sign}(\lambda v)\over \sqrt{\xi^3}} 
\left |\tan\left({N\lambda\sqrt{\xi}\over 2}
\right)\right|\right],
\end{eqnarray}
\end{mathletters}
for $\xi>0$ and
\begin{mathletters}
\label{sol2}
\begin{eqnarray}
&&N={2\over |\lambda |\sqrt{\xi}}\log \left[{1+ \sqrt{2-\sqrt{4-\gamma\xi}}
\over 1-\sqrt{2-\sqrt{4-\gamma\xi}}}\right ]\\
&&E={m v^2\over 2|\lambda|}\left[{(1+|\xi|)\over |\xi|}{N|\lambda|}-
{2~\over \sqrt{|\xi|^3}}
\tanh\left({N |\lambda|\sqrt{|\xi|}
\over 2}\right)\right],
\end{eqnarray}
\end{mathletters}
for $\xi<0$. We observe that for $\xi>0$ $N$ takes value in a closed interval:
\begin{mathletters}
\begin{eqnarray}
0\le &N& \le {\pi\over |\lambda| \sqrt{\xi}}\,\,\,\,\,\,\,
\lambda v<0,\\
{\pi\over |\lambda| \sqrt{\xi}} 
\le &N& \le {\over }{2\pi\over |\lambda|\sqrt{\xi}}\,\,\,\,\,\,\,
\lambda v >0.
\label{Disse}
\end{eqnarray}
\end{mathletters}
No restriction arises  for $\xi<0$. The formulae for the 
energy and the number present a potential singularity at $\xi=0$; actually 
for ${\rm sign}(\lambda v)<0$ the limit can be taken, recovering the previous 
results and the expressions in the two ranges are connected by  analytic
continuation. It is worth noticing that the energy is a non-polynomial 
function of $N$ and $\lambda$ (except for $\xi=0$), but an effective mass 
can be introduced as well, in perfect analogy with the zero potential case. 

Finally we observe that for all these solutions (for any $\xi$) the sign of
the total momentum $P$ is opposite to that of the  coupling constant
$\lambda$, in spite of the fact that for $\xi>0$ we can have  negative 
and positive velocity (the total energy has the correct sign in order to 
realize the described situation). Chirality (in this more general meaning,
{\it i.e.}, $\lambda P<0$) 
is therefore a property shared by these scale invariant systems and it is 
clearly related to the non-linear current interaction present in the relevant 
Schr\"odinger equation. At the quantum level we shall recover such a property in the 
two-body quantum bound state and we shall  make some observations for the 
general case.

\subsection{Solutions around the non-trivial vacuum}

We investigate traveling solutions  ${\phi}(x,t)$ that  approach, at spatial
infinity,  a vacuum with  density $\rho_{0}\ne 0$. When interpreting 
the  standard non-linear Schr\"odinger equation as a description of a gas of bosons 
interacting via a repulsive two-body potential, this boundary condition is often 
called the ``condensate boundary condition'', since a 
time-independent solution 
$\rho=\rho_{0}$ is usually understood as the Bose condensate. Finite 
energy solutions with the above boundary are therefore seen  as 
travelling bubbles in a constant background density. 
%In our non-galilean 
%invariant system one as to specify one more physical parameter, namely the 
%value of the current at $\infty$.
%
%\noindent
%The requirement that $\hat{\phi}(x,t)\rightarrow \hat{\phi}_{vacum}(x,t)$ 
%implies that $J_{0}=J(\pm \infty)=0$ fixing the arbitrary function of the 
%time appearing as integration constant
%\begin{equation}
%J_{\infty}(t)=-v\rho_{0}^2.
%\label{conditio}
%\end{equation}
On the other hand, the ``naive'' momentum $P$ is not  finite,  because of 
the presence of the ``unusual''  $\rho^4$ term in the density $T^{0}_{~x}$. 
In  the presence of non-vanishing boundary  conditions a  conserved charge 
involves contributions coming from the boundary as well. 
The correct definition of the total momentum requires the addition of a 
suitable boundary term, that makes $P$ differentiable (in the functional sense) 
and therefore compatible with the Hamiltonian structure of the system. The 
resulting momentum turns out to be finite. This procedure entails a breaking of 
the relation $E={v\over 2}P$, derived from the scale invariance, a fact that is 
nevertheless expected because our ``condensate boundary condition'' explicitly 
breaks such a symmetry. 

\noindent
We start by presenting the phase of $\hat{\phi}$ when  the non trivial boundary
condition  is  taken into account:
\begin{equation}
\theta(x,t)={m v\over \hbar}\int_{-\infty}^{x-vt}(1-{\rho_0^2\over \rho^2})
-\frac{\lambda}{2}\int_{-\infty}^{x-vt}(\rho^2-{\rho_0^2}) -
\frac{\lambda}{2}\rho_0^2 x+\theta_0
\label{fase}
\end{equation}
and obviously it approaches to the phase of the vacuum as $x\rightarrow 
\infty$. Then using eq.(\ref{fase}) we obtain the equation for $\rho(y)$ 
\begin{equation}
\rho^{\prime\prime}+\frac{2 m}{\hbar}\left ({m v^2\over 2 \hbar}+
{\lambda v \rho_0^2}\right)\rho-{m^2 v^2 \over \hbar^2}{\rho^4_0\over
\rho^3}-{2 m  v\lambda\over \hbar}\rho^3=0
\label{modolo1}
\end{equation}
or equivalently (by using the boundary condition)
\begin{equation}
(\rho^{\prime})^2=\bigl(\rho^2-\rho_0^2\bigr)^2
[{m v\over\hbar}\lambda-{m^2 v^2\over \hbar^2\rho^2}].
\label{modolo2}
\end{equation}
The relevant solution approaching to $\rho_0^2$ as $y\rightarrow\infty$ exists 
if and only if 
\begin{equation}
\label{condi3}
\lambda v >0 \ \ \ \ \ \ \ {\rm and}\ \ \ \ \ \ \
 \rho^2 >\frac{ m v}{\hbar\lambda}
\end{equation}
The shape of our ``dark soliton'' is
\begin{equation}
\label{modolo3}
\rho^2(x-vt)=\rho_0^2\left[1-\left (1-{m v\over\lambda\hbar \rho_0^2}
\right ){\rm sech}^2\left (
\sqrt{{ m v\lambda \rho_0^2\over \hbar} \left (1-{ m v\over
\hbar\lambda\rho_0^2}\right )}(x-vt)\right )\right].
\end{equation}
The dark soliton is chiral 
($\lambda v >0$) and moving in the opposite direction with respect to 
the soliton found in the previous section for vanishing cubic potential.
Moreover only a finite range of velocity is permitted as 
clearly displayed by eq.(\ref{condi3}): at variance with the soliton case we 
see that the only free parameter not 
fixed by the boundary conditions is the velocity $v$ (we keep
$\rho_0^2$ fixed); its absolute value cannot exceed the critical bound 
$v_{max.}=\displaystyle{\frac{\hbar\lambda\rho_0^2}{m}}$. The vacuum solution 
is recovered as $v$ approaches $v_{max.}$. 

Let us discuss the conserved quantities: the number, 
as defined in the previous 
section, is obviously not finite. It is natural, in the spirit of a bubble 
interpretation, to call number the quantity
\begin{equation}
N=\int_{-\infty}^{+\infty}dx\,(\rho_0^2-\rho^2),     
\label{numbero}
\end{equation}
that evaluated for eq.(\ref{modolo3}) is
\begin{equation}
N={2\over \lambda}\sqrt{{v_{max.}\over v}-1}.     
\label{numnero}
\end{equation}
$N$ is a monotonically decreasing function of $v$ for any choice of background 
density and coupling constant. Next the expression for the energy is
\begin{equation}
E={2\over 3}\hbar\rho_0^2\sqrt{{v\over v_{max.}}
\left(1-{v\over v_{max.}}\right )^{{3}}}.
\label{Energa}
\end{equation}
The naive momentum 
$\displaystyle{P=\int_{-\infty}^{+\infty}\left 
[mJ+\hbar{\lambda\over 2}\rho^4\right ]}$
is infinite, as we have anticipated,  because of the presence of the 
``non-galilean'' term (we recall that $J\rightarrow 0$ at infinity).
However 
our Schr\"odinger equation is a Hamiltonian system with phase space consisting 
of pairs of functions $({\phi}(x),{\phi}^*(x))$ with boundary condition 
$({\phi}(x){\phi}^*(x))\rightarrow\rho_0^2$ and the Poisson bracket
\begin{equation}
\{A,B\}=-i\hbar\int_{-\infty}^{+\infty}d\,x\left({\delta A\over\delta {\phi}(x)}
{\delta B\over\delta{\phi}^*(x)}-{\delta A\over\delta{\phi}^*(x)}
{\delta B\over\delta{\phi}(x)}\right).
\end{equation}
To compute the Poisson bracket of $P$ with some other functional we have to 
compute the functional derivatives $\displaystyle{
{\delta P\over\delta{\phi}(x)}}$ and $\displaystyle{
{\delta P\over\delta{\phi}^*(x)}}$: using the expression of $P$ as function 
the canonical pair we have
\begin{equation}
\delta P =i \frac{\hbar}{2}\rho_0^2 \left({\phi}^{*}
\delta{\phi}-{\phi}\delta{\phi}^*\right)\Biggr|^{+\infty}_{-\infty}
+i\hbar\int_{-\infty}^{+\infty}d\,x\left(\delta{\phi}\partial_x{\phi}^*
-\delta{\phi}^*\partial_x{\phi}\right).
\label{varitio}
\end{equation}
The functional derivative can be computed only if the first term on the
right does not 
appear, so we have to add a functional that cancels the boundary term: this 
functional is proportional to $\theta^{\prime}$ and the new momentum can be 
written in the compact form
\begin{equation}
\hat{P}={i\hbar\over 2}\int_{-\infty}^{+\infty}d\,x\left({\phi}^{*}
\partial_x{\phi}-{\phi}\partial_x{\phi}^{*}\right)
\left(1-{\rho_0^2\over 
{\phi}{\phi}^{*}}\right).
\label{Momnto}
\end{equation} 
$\hat{P}$ is automatically finite and obviously the subtraction is 
equivalent to a suitable improvement in the energy-momentum tensor. We 
define the improved components
\begin{equation}
\hat{T}^{0}_{~x}=T^{0}_{~x}-\displaystyle{{i\hbar \rho_0^2\over 2}}
\partial_{x}\ln\left({{\phi}\over {\phi}^*}\right),\ \ \  \  \
\hat{T}^{x}_{~x}=T^{x}_{x}+
\displaystyle{{i\hbar \rho_0^2\over 2}}
\partial_{t}\ln\left({{\phi}\over {\phi}^*}\right)
\label{eq:improved2}
\end{equation}
automatically satisfying the conservation equation and giving
$\displaystyle{
\hat{P}=\int_{-\infty}^{+\infty}d\,x\hat{T}^{0}_{~x}.
}$
We remark that even the integral of $\hat{T}^{x}_{~x}$ is convergent.
A further observation concerns the scale invariance of the theory in presence 
of the condensate boundary condition: the generator $D$ 
(that is not defined using the naive expression for $P$) is not conserved, as 
we can expect
\begin{equation}
{dD\over dt}=-\displaystyle{{i\hbar \rho_0^2\over 2}}\int^\infty_{-\infty} dx
\partial_{t}\ln\left({{\phi}\over {\phi}^*}\right)
%{\rho_0^2\over 4}\int_{-\infty}^{+\infty}d\,x
%\bigl[{\hbar^2\over 2m}
%{\hat{\phi}^*(\partial_x+i{\lambda\over 2}\rho^2)^2\hat{\phi}+
%{\hat{\phi}(\partial_x-i{\lambda\over 2}\rho^2)^2\hat{\phi}^*
%\over \hat{\phi}^*\hat{\phi}}}+\hbar\lambda J\bigr].
\label{nonconse}
\end{equation}
The energy-momentum relation is therefore modified as follows
\begin{equation}
H={v\over 2}\hat{P}+{dD\over dt}.
\label{newrella}
\end{equation}
The momentum $\hat{P}$ of the dark soliton is easily evaluated
\begin{equation}
\hat{P}=-{\rm sign}(\lambda)\hbar\rho_0^2
\left[2{\rm arctg}
\sqrt{{v_{max.}\over v}-1}
+{4\over3}\left({v\over v_{max.}}\right)
\sqrt{\left( {v_{max.}\over v}-1\right)^{3}}-
\sqrt{{v_{max.}\over v}-1}
\right],
\label{totalmom}
\end{equation}
that has been checked to be consistent with eq. (\ref{newrella}). We notice 
that $\hat{P}$ is a monotonically decreasing function of $v$, ranging from 
$+\infty$ to $0$ (in absolute value), while the energy has a maximum value 
$E=\hbar \displaystyle{
{\sqrt{3}\over 4|\lambda|} \left({m v_{max.}^2\over 2} \right)}$ for 
$\displaystyle{v={v_{max.}\over 4}}$. It is 
interesting to obtain the dispersion relation for these solutions, namely 
$\displaystyle{{dE\over d\hat{P}}}$. In a particle-like solution we expect
$\displaystyle{
{dE\over d\hat{P}}=v.}
$
In ref.\cite{Russi} the same subtraction procedure was followed in a Galilean 
invariant theory for dark soliton solutions: in that case the particle-like 
dispersion relation was recovered. This is not true in our case: performing the 
variation of the energy respect $\hat{\phi}^*$ and $\hat{\phi}$ 
and using the equation of motion we get
\begin{equation}
\delta E=v\delta\hat{P}-v{\hbar \lambda\over 2}\rho_0^2\delta N,
\end{equation}
and therefore 
\begin{equation}
{dE\over dv}=v{d\hat{P}\over dv}-v{\hbar \lambda\over 2}\rho_0^2{dN\over dv}.
\label{disperso}
\end{equation}
Because the theory is not Galilean invariant the number depends on $v$, as we 
have seen, once $\rho_0^2$ is considered a fixed parameter: the explicit 
computation gives
\begin{equation}
{dE\over d\hat{P}}=2v{(v_{max.}-v)(v_{max.}-4 v)\over 8v^2-10 v v_{max.}-
v^{2}_{max.}},
\label{dispersio2}
\end{equation}
which does not resemble the particle-like behavior.
The curve $E(\hat{P})$ can be easily plotted
\begin{figure}
\vskip -1truecm
\hskip
2truecm\epsfysize=7truecm\epsfxsize=11truecm\epsfbox{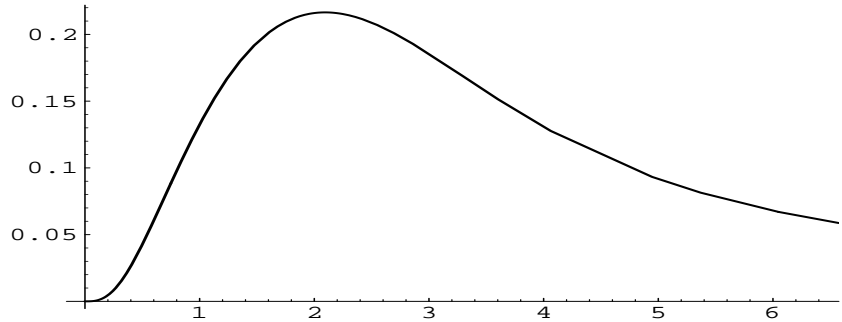}
\caption{The x-axis is $-{\rm sign}(\lambda)
 ~ \displaystyle{\frac{P}{\hbar\rho_0^2}}$,
  the y-axis is $\displaystyle{\frac{E}{\hbar\rho_0^2}}$.}
\end{figure}
\noindent
The total energy goes to zero for small and large momenta and in particular 
as $\hat{P}\rightarrow 0$ we have
\begin{equation}
E\simeq {2\over 3}{|\lambda|\over \hbar m \rho_0^2}|\hat{P}|^3. 
\end{equation}
We have seen that the finite-energy condition is equivalent to the
requirement 
that $J(\pm\infty)=0$. We can also imagine physical situations in which the 
asymptotic current is not zero; in this case a subtraction procedure can 
derived from the differentiability requirement, leading to a sensible
definition of the total energy. The general structure of the dark solitons in this case 
will be reported elsewhere.

We turn now, as in the previous subsection, to the problem of generalizing the 
above analysis when a cubic potential in $\rho^2$ is added. A finite energy 
configuration requires that asymptotically $\hat{\phi}$ approaches  a 
static solution with constant modulus and zero energy: the phase of these waves 
must satisfy
\begin{equation}
(\theta^{\prime}+{\lambda\over 2}\rho_0^2)^2={\lambda^2\over 4}\xi\rho_0^4.
\label{wowa}
\end{equation} 
Therefore only for $\xi\geq 0$  can we obtain the static, zero-energy solution
\begin{equation}
\theta=-\frac{\lambda}{2}\rho_0^2\left (1\mp \sqrt{\xi}
\right )x+\theta_0
%({\beta\over 4\alpha}\rho_0^2)^2\pm\sqrt{{\beta^2\xi\over 16\alpha^2}
%})\rho_0^2x+\theta_{0}(t),
\label{wowa1}
\end{equation} 
that carries the asymptotic current
$\displaystyle{
J=\pm \frac{\hbar \lambda}{2 m}\sqrt{\xi}\rho_0^4}.$
%\sqrt{{\beta^2\xi\over 4}}\rho_0^2.                 
When considering dark solitons interpolating these solutions at $\pm\infty$, 
this asymptotics  reflects  the following choice for $J_{\infty}$:
\begin{equation}
J_{\infty}(t)=-v\rho_0^2\pm \frac{\hbar \lambda}{2 m}\sqrt{\xi}\rho_0^4.
\end{equation}
The relevant equation for the phase is easily solved
\begin{equation}
\theta=\left({m v\over \hbar}\mp {\lambda\over 2 }\sqrt{\xi}\rho^2_0\right)
\int_{-\infty}^{x-vt}\!\!\!\!\!\!d\,y \left (1-{\rho_0^2\over \rho^2}\right)-
{\lambda\over 2}
\int_{-\infty}^{x-vt}\!\!\!\!\!\!d\,y(\rho^2-\rho_0^2)-
\frac{\lambda}{2}\rho_0^2\left (1\mp\sqrt{\xi}\right ) x +\theta_0(t),
\end{equation}
while for the modulus we get
\begin{equation}
(\rho^{\prime})^2=(\rho^2-\rho_0^2)^2\left [ \frac{m v}{\hbar}\lambda-
\left(\frac{m v}{\hbar}\mp\frac{\lambda\sqrt{\xi}}{2}\rho_0^2\right)^2
\frac{1}{\rho^2}-\frac{\lambda^2}{4}\xi (\rho^2+2\rho_0^2)\right]
%[-{(2v\mp\sqrt{\beta^2\xi\rho_0^2})^2\over 16}\alpha^2{1\over \rho^2}-
%({2\beta v+\beta^2\xi\rho_0^2\over 8\alpha^2})
%-{\beta^2\xi\over 16\alpha^2}\rho^2]. 
\end{equation}
From the analysis of the above equation we can derive the 
parameter range for which solitonic solution exist: one has to solve a 
system of inequalities and here we state only the result, deferring details 
and physical interpretation to a forthcoming paper focused on "condensate 
boundary condition" solutions.  

\noindent
$a.)\,\,\,\,\,$ The existing solutions are chiral with sign$(\lambda v)>0$;

\noindent
$b.)\,\,\,\,\,$ There is an intrinsic lowest velocity 
$$|v|>{4\over 3}{\hbar|\lambda|\over m}\xi\rho_0^2$$

\noindent
There is no solution with finite energy unless $0\leq\xi\leq 1$; 

\noindent
In this range we have two different kinds of solutions: setting
\begin{eqnarray}
A&=&-{\beta v\over 4\alpha^2}-{3\beta^2\xi\rho_0^2\over 16\alpha^2}
-{(2v\mp\sqrt{\beta^2\xi\rho_0^2})^2\over 16\alpha^2\rho_0^2},\nonumber\\
B&=&{\beta v\over 4\alpha^2}+{\beta^2\xi\rho_0^2\over 4\alpha^2},\nonumber\\
C&=&-{\beta^2\xi\rho_0^2\over 16\alpha^2},\nonumber\\
\Delta&=&B^2-4AC,\nonumber\\
\end{eqnarray}
with $\alpha=\displaystyle{\frac{\hbar}{2 m}}$ and
$\beta=-\displaystyle{\frac{\hbar\lambda}{2m}}$ we find a true dark soliton
\begin{equation}
\rho^2=\rho_0^2\bigl(1-{2A\over \sqrt{\Delta}\cosh(2\rho_0\sqrt{A}(x-vt)) -B}
\bigr),
\end{equation}
that represents an ``hole'' in the background density and a ``bright'' soliton
\begin{equation}
\rho^2=\rho_0^2\bigl(1+{2A\over \sqrt{\Delta}\cosh(2\rho_0\sqrt{A}(x-vt))+B}
\bigr),
\end{equation}
that is a positive density excitation over the background. Obviously while the 
first one goes smoothly to the solution for $\xi=0$ the second one has a 
singular limit. The upper limit on the velocity is determined by the 
positivity of $\Delta$.

We end this section by remarking that ``condensate boundary condition'' 
solutions have no  clear interpretation in the quantum theory; although their 
quantum role is an open problem they find applications in such branches 
of physics as condensed matter and plasma dynamics.

\section{Quantum theory}
%\baselineskip .8truecm

The simplest non relativistic example of the relation between solitons
and  quantum bound states is represented by the usual nonlinear 
Schr\"odinger equation. One may view eq. (\ref{eq:10}) as a Heisenberg
equation for the quantum field $\hat\phi(x,t)$,  which is taken to satisfy
the commutation relation
\begin{equation}
[\hat\phi(x_1,t),\hat\phi^{*}(x_2,t)]=\delta(x_1-x_2).
\label{commo}
\end{equation}
The Hilbert space can be decomposed into invariant subspaces according to
the integer eigenvalues $n$ of the (conserved) number operator
$\displaystyle{\int_{-\infty}^\infty dx\,\hat\phi^{*}\hat\phi}$ and 
the $n$-body 
wave function,
\begin{equation}
\phi_{n}(x_1,...,x_n,t)={1\over\sqrt{n}}<0|\hat\phi(x_1,t)...\hat\phi(x_n,t)|n>,
\label{vave}
\end{equation}
satisfies a Schr\"odinger equation with two-body, pairwise  attractive
$\delta$-function interactions. The bound-state spectrum can be explicitly
computed \cite{MAC} and in the center of mass frame it is given by
\begin{equation}
E_{n}=-{\lambda^2m\over 6\hbar^2}(n^3-n).
\label{spettro}
\end{equation}
On the other hand, the semiclassical quantization of the one soliton
solution at the leading order produces the identification of the classical
$U(1)$ charge $N=\displaystyle{\int^{+\infty}_{-\infty} dx~ \hat\phi_{soliton}^*\hat\phi_{soliton}}$
with the quantum number $n$ and  the spectrum
%The static one soliton solution of the classical nonlinear equation has energy 
%$$E_{Cl.}(N)=-{\lambda^2m\over 6\hbar^2}N^3$$, where $N$ is the value of 
%$\int_{-\infty}^{+\infty}dx\hat\phi^{*}\hat\phi$ evaluated on the solution. 
%The semiclassical quantization of the soliton can be performed leading to
\begin{equation}
E_{Semicl.}(n)=E_{Class.}(n)=-{\lambda^2m\over 6\hbar^2}n^3.
\end{equation}
Note that the classical energy of the soliton $E_{Class.}$ gives the
correct leading term as $\lambda\to 0$ and $\lambda n $ is kept fixed.
Remarkably the next-to-leading correction diminishes $n^3$ by $n$
($n^3\to n^3-n$) and it is exact. From the quantum field theory point
of view the correction linear in $n$ to the energy stems from a renormalization
effect, namely from the counterterm $m(\Lambda)\hat\phi^*\hat\phi$ necessary to cancel 
the divergences of the theory.
%
%a counter term of ther form $m(\Lambda)\psi^*\psi$ must be introduced
%the first quantum correction diminishes $N^3$ by $N$ (so that $E_1$ is the 
%energy of the free particle) and it is exact. It is remarkable to
%
%the classical energy gives the leading term as  $\lambda\rightarrow 0$, $N\lambda={\it cost.}$: moreover the quantum correction consists only of a 
%renormalization term in the effective action ($\Lambda\psi^*\psi$) \cite{NOL}. 
%The quantum energy coincides with classical energy of the theory where the 
%above term is included with the normalization condition that $E_1$ is the free 
%particle:

One can argue on general ground \cite{FADE} and verify by
computation \cite{DHN,COREA,NOL} that all integrable systems enjoy a similar 
stringent analogy between quantum bound states and solitons; much 
less is known for nonintegrable  
ones\footnote{From  the point of view of the form factor approach,
the question was recently investigated  in \cite{PUCCI}, where a 
a non-integrable perturbation of the Ising model in external magnetic 
field was considered.}. Our model ($\xi=0$)  appears to belong to this
second category: indeed it fails  to pass the Painlev\'e test.  
Nevertheless it can be considered as a perturbed version of the 
derivative nonlinear Schr\"odinger equation ($\xi=1$)\cite{DAS}, 
for which  some evidences of quantum integrability \cite{INDIA} 
exist. Thus it seems reasonable to expect that the quantization 
of the corresponding one-soliton solution will provide, in this 
case,  a good approximation of the quantum spectrum.  However 
before committing ourselves with the semiclassical approach, we
shall try to extract some information through the pattern described 
at the beginning of this section.

We take for the quantum hamiltonian the normal ordered expression 
\begin{equation}
H={\hbar^2\over 2m}\int dx: \left|\left(\partial_{x}+i
{\lambda\over 2}\rho^2\right){\phi}\right|^2:
\label{Hammo}
\end{equation}
and we posit the canonical commutation relations (\ref{commo}). The normal 
ordering prescription has been adopted to define properly the quantum  
Hamiltonian. This corresponds to removing all the singular 
interactions proportional to $\delta(0)$ in the resulting $n$-body Schr\"odinger 
equations. The above procedure is not without price, entailing in fact the 
loss of positivity for the quantum energy. 
[ The classical Hamiltonian  \underbar{is} positive definite for ${\it
any}$ value 
of the coupling  constant $\lambda$.] With this choice, the $n$-body 
Schr\"odinger equation becomes
\begin{equation}
-{\hbar^2\over 2m}\left [\sum_{i=1}^{n}\partial^{2}_{x_i}+2i\lambda
\sum_{i<j}^{n}\delta(x_i-x_j)\partial_{x_i}-
{\lambda^2\over 4}\sum_{{i\neq j\neq k}}^{n}
\delta(x_i-x_j)\delta(x_i-x_k)\right]
\phi_n=E_n\phi_n.
\label{eigen1}
\end{equation}
In eq. (\ref{eigen1}) the quantum non-integrability manifests itself 
through the presence of a three-body interaction generated by the 
$(\phi^*\phi)^3$ part of the Hamiltonian. In the case of the 
DNLS equation ($\xi=1$) this term is, instead, cancelled by an 
analogous  contribution coming from the potential $V$.
By translational invariance we can always separate the center 
of mass motion by introducing 
the parametrization 
\begin{equation}
\phi_n(x_1,..x_n)=\exp\left[i\sum_{i=1}^{n}{Px_i\over n}\right]\chi_n(x_1,..x_n),
\ \ \ \ {\rm with}\ \ \ \sum_{i=1}^{n}\partial_{x_i}\chi_n(x_1,..x_n)=0,
\label{para}
\end{equation}
namely $\chi_n$ depends only on the relative coordinates. However, because
of the lack Galilean invariance, the total momentum $P$ will  not decouple 
completely and  will appear as a parameter in the reduced Schr\"odinger 
equation for $\chi_n$.

From (\ref{eigen1}) we note that the  claim in \cite{RABE} of
statistical transmutation for the quantum  excitation of this model 
is inexact. It is impossible to remove the $\delta$ interaction
by a phase redefinition and therefore no change in the statistical behaviour 
should be expected \cite{NOI}.

\subsection{The two-body problem}

We begin our quantum investigations by considering the simplest sector 
contained in the Hilbert space of the field theory defined by the quantum 
Hamiltonian (\ref{Hammo}): the two-particle one. Using eq. (\ref{para}) 
and defining $x\equiv x_1-x_2$ we find that $\chi_2$ satisfies
\begin{equation}
\left(-{\hbar^2\over m}{\partial^2\over\partial x^{2}}+{P^2\over 4m}+
\hbar{\lambda P\over 2m}\delta(x)\right)\chi_2(x)=E_2 \chi_2(x).
\label{sepae}
\end{equation}
The presence of the total momentum $P$ in the $\delta-$function potential 
for the relative motion vividly demonstrates the absence of Galilean
invariance. Provided that
\begin{equation}
\lambda P<0
\label{chiral}
\end{equation}
eq.~(\ref{sepae}) possesses a bound state solution with energy
\begin{equation}
E_2={P^2\over 4m}\left(1-{\lambda^2\over 4}\right)=\frac{P^2}{2 M_{B.S.}}.
\label{legae}
\end{equation}
Since $P/2 m$ can be identified with the velocity $v$ of the center
of mass, we recognize the condition (\ref{chiral}) as the one (see eq.
(\ref{condi})) that guarantees the existence of the soliton (\ref{prisoli})
at the classical level. Remarkably even the scale invariance is preserved:
in fact the dispersion relation (\ref{legae}) is unchanged under dilatation.
[Unfortunately this property will fail in the $n$-body problem.] On the 
other hand  the energy $E_2$ and therefore the effective mass of the bound
state
\begin{equation}
M_{B.S.}={2m\over\displaystyle{ 1-{\lambda^2/4}}}
\end{equation} 
become negative for $\lambda^2>4$. The loss of positivity originates from
the normal ordering adopted in defining the quantum Hamiltonian. As a 
three-body interaction of the form $(\phi^*\phi)^3$ can never contribute
to the two body problem, the bound state (\ref{legae}) is, actually, shared
by all the family (\ref{models}) of models defined in sect. 2 and 
parametrized by 
the coupling constant $\xi$. Moreover all the solitons in this class of
models enjoy the property $\lambda P<0$ in perfect analogy with eq.
(\ref{chiral}).

It is instructive to recover the same results in a  field  theory framework, 
namely by resumming the  perturbative series defined in terms of Feynman 
graphs. This approach will be most useful in the discussion of the 
$n-$body case. The quantum propagator  for $\phi$ is easily derived 
\begin{equation}
D(x,t)={1\over (2\pi)^2}\int dk~d\omega~{e^{-i(\omega t-kx)}i\over 
\omega-{k^2\over 2m}+i\epsilon},
\label{propa}
\end{equation}
($\hbar$ is put equal 1 from now on) and the interaction vertices are
\begin{figure}
%\vskip 4truecm
%\epsfysize=6truecm
\hskip 2truecm\epsfbox{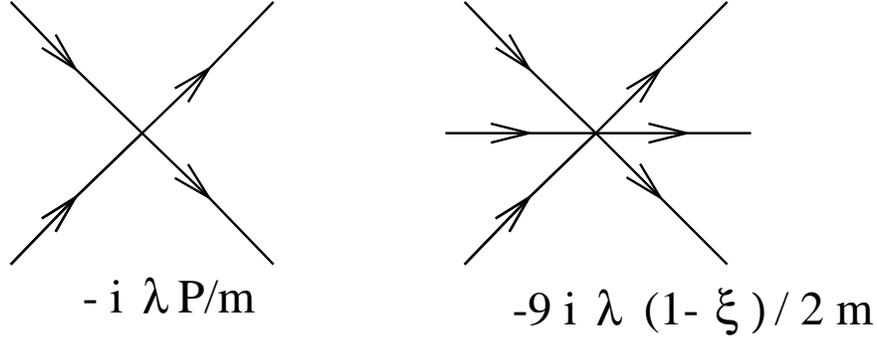}
%\smallskip
\caption{Feynman rules: P is the total incoming momentum.}
\end{figure}
\noindent
At one-loop level the relevant  graph is 
\begin{figure}
%\vskip 4truecm
%\epsfysize=5.5truecm
\hskip 2truecm\epsfbox{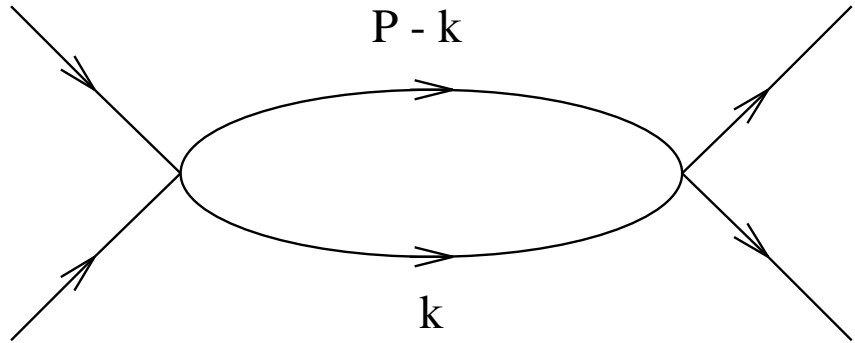}
\medskip
\caption{One loop contribution to the four point vertex. P is the
total incoming momentum, k the loop momentum.}
\end{figure}
\noindent
The absence of the $t$ and $u$ channel implied by the non-relativistic 
character of the theory and  moreover the absence of a tadpole contribution,
because of the normal-ordering, greatly simplifies the final result. Choosing 
\footnote{This restriction  is  the equivalent of the condition 
$\gamma>0$ found in the discussion of the classical soliton in subsect. 3A}
$P^2>4mE$,
we get the one-loop amplitude
\begin{equation}
A_{1}(P,E,\lambda)=-i{\lambda^2 P^2\over 2m}{1\over \sqrt{P^2-4mE}}.
\label{ampli1}
\end{equation}
Here $P=p_1+p_2$ is total momentum and $E=\omega_1+\omega_2$ is the 
total energy. To obtain the full amplitude we have only to resum the 
geometric series
\begin{figure}
%\vskip 4truecm
\hskip .5truecm \epsfbox{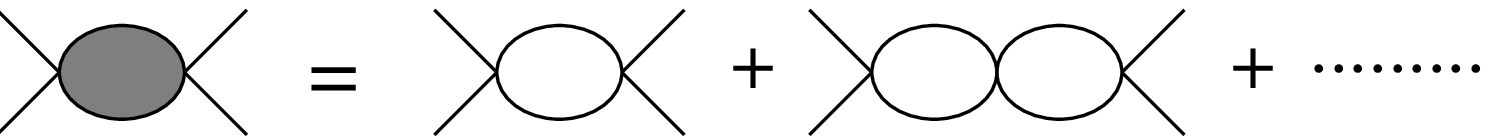}
\end{figure}
\noindent
because there is no contribution  from higher loops or from the six-vertex.
[Recall in fact that the number of particles must be conserved in each
stage of our diagrams since we are in a non-relativistic theory.] Thus we 
obtain
\begin{equation}
A(P,E,\lambda)=i{\lambda P\over m}{\sqrt{P^2-4mE}\over 
{\lambda P/ 2}+\sqrt{P^2-4mE}}.
\label{ampli2}
\end{equation}
We have a pole for $\lambda P<0$ at the energy  $\displaystyle{
E={P^2\over 4m}
\left(1-{\lambda^2\over 4}\right)}$, recovering therefore 
the  results of the two-body Schr\"odinger equation. We remark that no 
infinity arises in the perturbative evaluation of the two-body scattering 
amplitude (once the normal-ordering is adopted) and no renormalization scale is 
needed: scale-invariance is maintained as the classical energy-momentum 
relation clearly shows.

\subsection{The n-body problem}

For the general $n$-body problem we did not succeed in computing  the 
energy eigenvalues for the bound states exactly because of the presence of the three-body 
interaction that complicates the eigenvalue equation. Nevertheless starting 
from the  usual non-linear Schr\"odinger equation    we can construct a 
perturbative solution to the problem, that is reasonable for certain values 
of the parameters and  is in agreement with the soliton characteristics.
The reduced equation is
\begin{eqnarray}
&&-\frac{1}{2m}\left[\sum_{i=1}^{n}\partial^{2}_{x_i}-
\frac{2\lambda P}{ n}
\sum_{i<j}^{n}\delta(x_i-x_j) +2i\lambda
\sum_{i<j}^{n}\delta(x_i-x_j)\partial_{x_i}-\right. \nonumber\\
&&\left. {\lambda^2\over 4}\sum_{i\neq j\neq k}^{n}
\delta(x_i-x_j)\delta(x_i-x_k)\right]
\chi_n=\left(E_n-\frac{P^2}{2 m n}\right)\chi_n.
\label{eigen3}
\end{eqnarray}
Defining $\displaystyle{z_i=-{2\lambda P\over n} x_i}$  we obtain
\begin{eqnarray}
&&\left[\underbrace{
\sum_{i=1}^{n}-\partial^2_{z_i}+{\rm sign}(\lambda P) \sum_{i<j}^{n}
\delta(z_i-z_j)}_{{\cal H}_{0}}-\lambda\underbrace{i
 {\rm sign}(\lambda P)\sum_{i<j}^{n}\delta(z_i-z_j)
\partial_{z_i}}_{{\cal H}_{1}}+\right.\nonumber\\
&&\left.{\lambda^2\over 4}\underbrace{
\sum_{i\neq j\neq k}^{n}\delta(z_i-z_j)\delta(z_i-z_k)}_{{\cal H}_{2}}
\right ]\chi_n
={m n^2\over 2 P^2\lambda^2}\left(E_n-{P^2\over 2 m n}\right)\chi_n.
\label{eigen4}
\end{eqnarray}
We recognize that the eigenvalue equation can be rewritten as the NLS one 
plus two perturbations proportional to $\lambda$ and $\lambda^2$
\begin{equation}
\left
[{\cal H}_0-\lambda{\cal H}_1+{\lambda^2\over 4}{\cal H}_2
\right]\chi={\cal E}_n \chi.
\label{perturba}
\end{equation}
The NLS equation defined by ${\cal H}_0$ admits bound states only if
$\lambda P<0$ in agreement with the chiral nature of the soliton; 
thus we shall consider only this case. The zero-order problem ${\cal H}_0 
\chi_n^{(0)}={\cal E}_n^{(0)} \chi_n^{(0)}$ gives
\begin{mathletters}
\begin{eqnarray}
&&{\cal E}_n^{(0)}=-{1\over 48}{(n^3-n)}\\
&&\chi_n^{(0)}=\sqrt{{[2(n-1)]!!\over 4^{n-2}n}}\exp\left
[-{1\over 4}\sum_{i\neq j} |z_i-z_j|\right],
\end{eqnarray}
\end{mathletters}
and therefore the ``zero-order'' energy
\begin{equation}
E_n^{(0)}={P^2\over 2 m n}\left (1-{n^2-1\over 12 }\lambda^2\right).
\end{equation}
This result is correct as $\lambda\rightarrow 0$,  
$\displaystyle{{P \lambda\over n}}$ 
is kept constant. Moreover the first correction, coming from  ${\cal H}_1$ 
vanishes. Infact if ${\cal E}_n={\cal E}_n^{(0)}+\lambda{\cal E}_n^{(1)}+..$ 
the  first order correction is simply
\begin{equation}
\!\!\!
{\cal E}_n^{(1)} \propto\!\!\!\int_{\infty}^{\infty}\Pi_{j=1}^{n}dz_j 
(\chi_n^{(0)})^{*}(\sum_{i<j}^{n}\delta(z_i-z_j)\partial_{z_i})\chi_n^{(0)}
\propto\!\!\!\int_{\infty}^{\infty}\Pi_{j=1}^{n }dz_j (\chi_n^{(0)})^{*}\chi_n^{(0)}
(\sum_{i\neq k}^{n}{\rm sign}(z_i-z_k))=0,
\end{equation}
because $(\sum_{i\neq k}^{n} {\rm sign}(z_i-z_k))=0$. Therefore
\begin{equation}
E_n={P^2\over 2m n}\left(1-{n^2-1\over 12 n}\lambda^2\right)+O(\lambda^4).
\end{equation}
Obviously for $n=2$ the result is exact: for generic $n$ it shows that in a 
suitable limit the equation is essentially equivalent to NLS equation with effective 
coupling constant $\displaystyle{{\lambda P\over n}}$. It is interesting to 
confront the perturbative solution with the exact energy spectrum of DNLS 
equation (${\cal H}_2=0$), which has been recently computed \cite{DAS}  
\begin{equation}
E_{n}^{DNLS}={P^2\over 4m}{\lambda\over 
\displaystyle{\tan\left(n\arctan{\lambda\over 2}\right)}}
\simeq {P^2\over 2m n}\left(1-{n^2-1\over 12 n}\lambda^2\right)+O(\lambda^4).
\end{equation}
The agreement between  the perturbative and the exact result provides a strong 
check of our approach.

\subsection{Trace anomaly}

In subsec. 4A we have seen how to recover the two-body states by means
of Feynman graphes. Here we would like to take advantage of the field theory
approach to derive another exact result (connected to scale invariance)
that would otherwise be quite difficult to establish. Classically the theory
is invariant under dilatation; at the quantum level, instead, the ultraviolet
divergences we encounter in perturbatively evaluating scattering amplitudes
will spoil this property. In fact the well-known machinery of renormalization
allows us to remove infinities consistently, but the price to be paid is the
introduction
of a ``renormalization scale'' that obviously breaks scale symmetry. In 
quantum mechanical language it is the highly singular three-body interaction
that needs to be defined: different choices correspond to different 
self-adjoint extensions of a Schr\"odinger operator associate with the
$n-$ body problem. The (dimensional) parameter describing the possible
extensions is usually related to the renormalized coupling constant
\cite{ALBE}.

A power counting analysis of the relevant diagrams shows that all divergences
arise from the two loop  six-point function or from its iterations (see
figure below). No infinity comes from the 4-point vertex. 
\begin{figure}
\label{sixver}
%\vskip 4truecm
%\epsfysize=5truecm
\hskip 2truecm\epsfbox{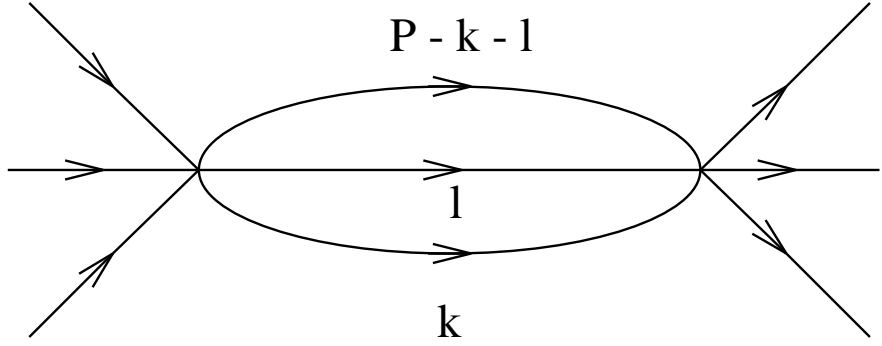}
\bigskip
\caption{Divergent 2-loop contribution to the six vertex function.
P is the total incoming momentum; k and l are the loop momenta.}
\end{figure}
\noindent
The two-loop amplitude
in fig. 4.  can be easily written down with the help of the
Feynman rules given in sect. 4A. After integrating over the energy
$\omega_1$ and $\omega_2$ in the loops we are left with the following
integral over the momenta $\left(\displaystyle{\alpha_0=
-\frac{9}{2 m}\lambda^2}(1-\xi_0)\right)$
\begin{eqnarray}
\label{pipina}
A_6(E,P)&&=-{i\alpha_0^2\over 24\pi^2}\int_{-\infty}^{+\infty}dl dk~
{1\over\displaystyle{
E-{l^2\over 2m}-{k^2\over 2m}-{(P-l-k)^2\over 2m}+i\epsilon}}
\nonumber\\
&&={i\alpha_0^2 m\over 12\sqrt{3}\pi^2}\int_{-\infty}^{+\infty}dk~ dl~
{1\over l^2 +k^2+\left(\displaystyle{{P^2\over 6}}-m E\right)-i\epsilon}.
\end{eqnarray}
Here, as before, $E$ and $P$ stand for the total energy and momentum
respectively. [Although we are mainly interested in $\xi=0$ case, we
are forced to introduce a bare coupling constant $\xi_0\ne 0$ in order
to have a renormalizable theory.] The final integral in
eq. (\ref{pipina}) is  logarithmically divergent. Upon introducing a 
cutoff $\Lambda$ and a  scale $\mu$, we can identify the singular 
contribution
\begin{equation}
A^{div.}_6(E,P)=
{i\alpha_0^2 m\over 12\sqrt{3}\pi}\log{\Lambda^2\over \mu^2}.
\end{equation}
The divergent term can be reabsorbed in a redefinition of the coupling
constant $\xi$ by defining a $\xi_R$ as follows
\begin{equation}
\xi_0=\xi_R-\frac{\sqrt{3}}{8\pi}\lambda^2 (1-\xi_R)^2 \log
\frac{\Lambda^2}{\mu^2}.
\label{renno}
\end{equation}
Since all the divergencies come from iterating the elementary graph 
in fig. 4 ( recall, in fact, that insertions of the four-point vertex 
do not give additional  contributions to the divergent part), the
result can be therefore made exact
\begin{equation}
1-\xi_0={1-\xi_R\over 1+\displaystyle{{\sqrt{3}\over 8\pi}}\lambda^2(1-\xi_R)
\log\displaystyle{{\Lambda^2\over \mu^2}}}.
\end{equation}
The $\beta$-function obtained from this relation is
\begin{equation}
\beta(\xi)=\mu{\partial\over\partial\mu}\xi_R(\mu)=
-{\sqrt{3}\over 4 \pi}\lambda^2(1-\xi_R)^2.
\label{betta}
\end{equation}
We remark that this result is exact and some of its consequences can 
be discussed. Firstly the coefficient of the three-body interaction
appears to be running (it depends on the scale $\mu^2$): in  field 
theoretical language it means that at the quantum level a potential
$V(\rho^2)\simeq \rho^6$ is induced from loop corrections. 
We are therefore naturally  led to consider the general family
parametrized  by $\xi$.  Secondly the fixed point of $\beta(\xi)$ is
for $\xi_R=1$, the DNLS equation, that is  the integrable system: scale
invariance is maintained in this model at  quantum level as the exact 
expression, {\it e.g.}, for the bound-state energy displays:
\begin{equation}
E={P^2\over 2M},\,\,\,\,\,\,\,\,\,M=m{\tan(N\arctan\lambda)\over \lambda}.
\label{arco}
\end{equation}
On the other hand for $\xi_R\neq 1$ we expect that the classical 
energy-momentum 
relation is broken at the quantum level. To make this idea more precise,
 we must
relate the scale dependence of the quantum theory to the non conservation 
of the generator $D$ of the dilatation. This can be done easily along the 
line of \cite{BARGA} where the same problem was discussed in 2+1 dimension for 
a $\rho^4$ theory: the methods relies in deriving a set of Ward
identities from 
scale invariance. Following \cite{BARGA} we obtain for the proper vertices in 
momentum-space $\Gamma^{n}(p_i,\omega_i,\xi_R,\mu)$ the Ward identity 
\begin{equation}
[\partial_a+(3-{n\over 2})]\Gamma^{n}(e^{-a}p_i,e^{-2a}\omega_i,\xi_R,\mu)=
-i\Gamma^{n}_{T}(0,e^{-a}p_i,e^{-2a}\omega_i,\xi_R,\mu),
\label{Ward1}
\end{equation}
$\Gamma^{n}_{T}(0,e^{-a}p_i,e^{-2a}\omega_i,\xi_R,\mu)$ being the proper 
vertex derived from the Green function
\begin{equation}
G^{n}_{T}(y,x_1,..x_n)=<0|T\bigl[\bigl(2 T^{00}(y)-T^{xx}(y)\bigr)\phi(x_1)...
\phi(x_n)\bigr]|0>,
\end{equation}
while $a$ is the dilatation parameter. If scale-invariance holds
 at the quantum 
level, the right-hand side of eq.(\ref{Ward1}) is zero. On the other hand 
the renormalization group equation, derived in the usual way is (we recall 
that no wave-function renormalization is present)
\begin{equation}
[{\partial \over \partial a}+\beta(\xi_R){\partial \over \partial\xi_R}+
(3-{n\over 2})]\Gamma^{n}(e^{-a}p_i,e^{-2a}\omega_i,\xi_R,\mu)=0.
\label{Ward2}
\end{equation}
Comparing eq.(\ref{Ward1}) and eq.(\ref{Ward2}) we see that scale invariance
is broken by the $\beta(\xi_R)$ term. More precisely we can write the 
anomalous Ward identity
\begin{equation}
\beta(\xi_R){\partial \over \partial\xi_R}\Gamma^{n}=i\Gamma_{T}^{n},
\label{Ward3}
\end{equation}
that reflects in a quantum equation for the trace of the energy-momentum tensor
\begin{equation}
T^{xx}=2T^{00}+{\sqrt{3}\over 32\pi}{\lambda^4\over m}(1-\xi_R)^2
(\phi^*\phi)^3.
\label{anoma}
\end{equation}
The scale anomaly is
\begin{equation}
{dD\over dt}=-{\sqrt{3}\over 64\pi}{\lambda^4\over m}(1-\xi_R)^2
\int_{-\infty}^{+\infty}(\phi^*\phi)^3.
\label{Anoma2}
\end{equation}
From the above analysis it appears clear that physical quantities, like the 
bound-states energies $E_n$, must depend both on $\xi_R(\mu)$ and
$\mu$ so that 
$\displaystyle{\mu {\partial\over \partial \mu}E_n(\xi,\mu)=0}$. For
example it is not 
difficult to compute the three-body bound-state in a simpler model, with 
potential purely proportional to $\rho^6$ 
($V(\rho^2)=-\displaystyle{{\alpha_0\over (3!)^2}\rho^6}$; the family of
models considered
here reduces to this one in the limit $\lambda\to 0$ but 
$\lambda^2 \xi_0=2 m \alpha_0/9$
is kept fixed).  In this case, due to the absence of the two-body
interaction,  we have only to resum the geometric series
\begin{figure}
%\vskip 4truecm
\hskip .5truecm\epsfbox{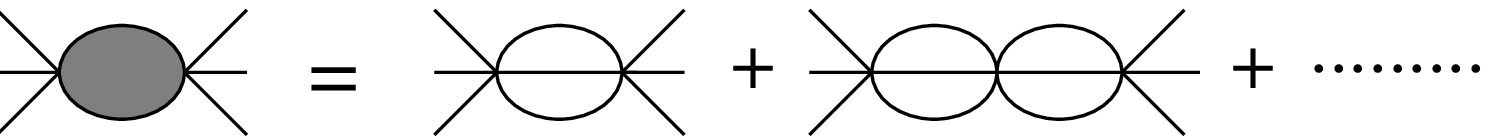}
\end{figure}
\noindent
leading to the exact scattering amplitude
\begin{equation}
A(P,E,\alpha)={i\alpha_R\over m}{1\over 1-
\displaystyle{\frac{\alpha_R}{12\sqrt{3}\pi}}\log
\displaystyle{{\mu^2\over P^2-6mE}}}
\end{equation}
that exhibits a pole for 
\begin{equation}
E_{3}={P^2\over 6m}-{\mu^2\over 6 m}\exp[-{12\sqrt{3}\pi\over
\alpha_R(\mu^2)}].
\end{equation}
The second term, that is actually independent from $\mu^2$
because the dependence of $\alpha_R$ on $\mu^2$ is so chosen
that this happens,  is generated by 
the scale anomaly and it breaks the classical energy-momentum relation. In 
term of the physical parameters the scattering amplitude is 
\begin{equation}
A(P,E,E_3)={12\sqrt{3}\pi i\over m ~\displaystyle{\log\left[{P^2-6E_3m\over P^2-6Em}\right]}},
\end{equation}
that depends on one more dimensionful parameter ($E_3$). Moreover we see that a 
finite energy bound-state requires the bare coupling constant $\alpha_0$ to be 
positive, implying that only the attractive potential leads to  non-trivial 
physics, in close analogy with the analysis performed in \cite{BARGA}. In this 
particular case the non-trivial interaction can also be viewed as a 
self-adjoint extension of the two-dimensional laplacian, with one point 
removed. The self-adjoint extension parameter has the interpretation of the 
renormalized coupling constant $\alpha$.

In our case the resummation, even in the simplest case of the three-body 
scattering amplitude, is not a geometric series: it is not clear if only a 
bare parameter $\xi_0>1$ leads to a bound-state or if the derivative coupling 
(for $\lambda P<0$) can bind particles even in presence of a bare repulsive 
three-body potential. In the next subsection we will try to combine the 
soliton structure of the theory with the trace anomaly in order to get some 
information on the quantum bound-state of the theory.

\subsection{Solitons and bound states}

The semiclassical quantization of the one-soliton solution needs essentially 
the computation of the small fluctuation around the soliton: following the 
approach of DHN \cite{DHN}, we have to calculate a functional determinant that 
produces the first quantum correction to the classical energy. In integrable 
systems, like NLS, DNLS or Sine-Gordon, one can use the multi-soliton
solutions 
to compute the stability angles \cite{DHN}, from which the determinant can be 
determined. In particular, in all these systems, the contribution of the 
determinant consists in a (finite) renormalization of the parameters 
appearing in the initial lagrangian. 
In the DLNS case, {\it e.g.}, the soliton has the 
mass $M_{Cl.}(N)=\displaystyle{{m\tan N\lambda\over \lambda}}$ while
the mass of the quantum  $n-$body bound-state is 
$M_{Q.}(n)=\displaystyle{{m\tan [n\arctan\lambda]\over \arctan \lambda}}$ \cite{DAS}.
The relevant renormalization is derived from the ratios
\begin{equation}
{M_{Cl.}(N)\over M_{Cl.}(1)}=\displaystyle{{\tan N\lambda\over
    \tan\lambda}},\ \ \ \ \ {\rm and}\ \ \ \ \ 
{M_{Q.}(n)\over M_{Q.}(n)}=\displaystyle{{\tan [n\arctan\lambda]\over
    \tan[\arctan\lambda]}},
\end{equation}
therefore consisting in the replacement $\lambda\rightarrow
\arctan\lambda$. [In
general we must expect that only the mass ratios are correctly given 
by the semiclassical approximation.] In Sine-Gordon we have a
similar  renormalization for the coupling constant $\beta^2$
$$
\beta^2\rightarrow\displaystyle{{\beta^2\over 
    1-\displaystyle{{\beta^2/ 8\pi}}}}.
$$
The identification of the number parameter $N$, with the quantum number $n$, 
representing the number of particles bounded can be obtained applying the 
Bohr-Sommerfield rule
\begin{equation}
dE=\omega dn,
\end{equation}
to the classical energy and taking $\omega$ the phase frequency at the 
maximum of the moving soliton (for $x-vt=0$).

In our case we do not have the two-soliton solution to use for the computation 
of the stability angles and it is not clear what is the structure of
the subleading  quantum corrections. We can nevertheless compute the
energy using the Bohr-Sommerfield formula, neglecting the contribution
of the functional determinant. This can be easily done for the general
family of scale invariant solitons, parametrized by $\xi$: we take the
energies of the classical soliton (\ref{sol1}) and (\ref{sol2}) and we
use the phase eq.(\ref{theta}), to obtain the frequency $\omega$ that
turns out to be
\begin{equation}
\omega=\omega_0+{m v^2\over\hbar}.
\end{equation}
We rewrite the Bohr-Sommerfield formula as 
\begin{equation}
{dE\over dN}=\omega{dn\over dN},
\end{equation}
that can be integrated to give $n=N$. The ``zero'' order energy for the n-body 
bound state is therefore
$$
E^{(0)}_{Q.}(n)=E_{Cl.}(n).
$$
Let us first study the case $\xi=0$; we expect that this equivalence is 
correct for $\lambda\rightarrow 0$. In fact
\begin{equation}
{E^{(0)}_{Q.}(n)\over E^{(0)}_{Q.}(1)}=
{1+\displaystyle{{\lambda^2 \over 12}}\over 1+
\displaystyle{{n^2\lambda^2\over 12}}}=
{1\over n}\bigl(1-{\lambda^2\over 12}(n^2-1)\bigr)+O(\lambda^4).
\label{cocca}
\end{equation}
We stress that this relation is valid to $O(\lambda^4)$ but in the case of
$n=2$ it
is exact (the quantum corrections must compensate the higher terms in the 
$\lambda$-expansion). In the general case we see that eq.(\ref{cocca}) is in 
agreement with the perturbative expansion developed in Subsect.3. However 
we have seen that from renormalization the general scale-invariant potential 
is induced and we are naturally led to the $\xi$-family of theories. We find 
that for $\xi>0$ different answers are given, in the small-$\lambda$ limit, 
depending on the soliton chosen:
\begin{eqnarray}
{E^{(0)}_{Q.}(n)\over E^{(0)}_{Q.}(1)}&=&
{1\over n}\bigl(1-{\lambda^2\over 12}(n^2-1)\bigr)+O(\lambda^4),
\,\,\,\,\, \lambda v<0, \nonumber\\
&=&{1\over n}\bigl(1+{\lambda^2\over 12}{\xi\over \xi-2}(n^2-1)\bigr)
+O(\lambda^4),\,\,\,\,\, \lambda v>0.
\end{eqnarray}
Only for $\xi=1$ do the two results coincide, while in general only for 
$\lambda v<0$ do we recover the ``perturbative'' computation: this is easily 
understood by realizing that only the soliton with $\lambda v<0$ is connected 
to an attractive NLS (that is the zero-order of the perturbative expansion). 
The meaning of the soliton with $\lambda v>0$ is not clear; moreover the entire
procedure, except that for $\xi=1$, can be trusted only for $P^2\rightarrow 
\infty$ and $\lambda\rightarrow 0$, because we are neglecting the
trace anomaly that drastically changes the momentum-energy relation
respected by our ``zero''-order approximation. Only in the large
momentum  limit does the classical term dominate over the ``anomalous'' 
contribution, as  can be inferred from the analysis of the pure
$\rho^6$ theory.

In the following, in order to improve our knowledge about the quantum
spectrum, we shall try to understand how the trace anomaly may modify 
the soliton analysis. We shall start assuming that the eigenvalue
$E_n$ has the form  
\begin{equation}
E_n=\frac{\mu^2}{m}F\left(\lambda,\xi_R(\mu^2), n,\frac{P^2}{\mu^2}\right)
\end{equation} 
The particular dependence on the mass $m$ can be inferred from the 
$n-$body Schrodinger equation. The factor $\mu^2$ has been pulled out 
to make the function $F$ dimensionless.  Since the energy levels are 
physical observables, they must be independent of the subtraction
point $\mu$, {\it i. e.} $\mu\displaystyle{\frac{\partial}{\partial\mu}} 
E_n(\mu)=0$, we have
\begin{equation}
t\frac{\partial F}{\partial t} 
-\frac{\beta(\xi)}{2} \frac{\partial F}{\partial \xi}- F  =0,
\end{equation}
where $t$ stands for $\displaystyle{\frac{P^2}{\mu^2}}$. It is not 
difficult to verify that the general solution for this equation is 
\begin{equation}
F= \frac{P^2}{\mu^2 {\cal M}\left [\lambda, n, 
\displaystyle{\frac{\mu^2}{P^2}} 
\exp\left({8\pi\over \sqrt{3}\lambda^2(1-\xi(\mu))}\right)\right]}
\end{equation}
or equivalently
\begin{equation}
\label{coppola7}
E_n= \frac{P^2}{ m {\cal M}\left [\lambda, n, 
\displaystyle{\frac{\mu^2}{P^2}}
\exp\left({8\pi\over \sqrt{3}\lambda^2(1-\xi(\mu))}\right)\right]}
\end{equation} 
The new dependence on the combination ${\mu^2}\exp\left({8\pi\over 
\sqrt{3}\lambda^2(1-\xi(\mu))}\right)$ is just the symptom of the 
appearance of a new scale in the theory. This, for example, may be
identified with the energy of the three-body bound state and it is 
obviously renormalization group invariant. 

In the limit $P^2\to\infty$, we expect a resurrection of scale
invariance and consequently eq. (\ref{coppola7}) must match, in
this limit, the soliton analysis. Taylor-expanding (\ref{coppola7})
we get 
\begin{equation}
\label{coppola8}
E_n= \frac{P^2}{ m {\cal M}_0\left [\lambda, n\right]}+ 
{\cal M}_1[n,\lambda]\mu^2
\exp\left({8\pi\over \sqrt{3}\lambda^2(1-\xi(\mu))}\right)
+ O\left(\frac{1}{P^2}\right).
\end{equation}
The first term (the dominant one in the infinite momentum limit)
respects scale invariance and we conjecture that the function 
${\cal M}_0[\lambda,n ]$ governing it  is the same as in the DNLS
equation. In fact we expect that such a model is the ultraviolet limit of 
our family of theories. An evidence of this is that $\xi_R(\mu)$ flows
to $1$ for $\xi_R (\mu_0)>1$. Notice that in a Galilean invariant theory 
correction of order $1/P^2$ are strictly forbidden, while they cannot
be excluded in principle here.
 
We conclude by observing that this brief analysis is consistent with
the Galilean invariant model $\rho^6$ considered at the end of the 
previous section.
 
\section{Conclusion}
In conclusion we have extensively studied a family of $1+1$ dimensional 
theories 
that describes non-relativistic bosons interacting with a gauge potential: the 
gauge action has been chosen to be of $B-F$ type plus a ``chiral'' kinetic 
term for $B$. This form was suggested from the dimensional reduction 
of Chern-Simons theory coupled to non-relativistic matter and it represents 
a simple way to introduce chiral excitations that can be important in some 
condensed matter context. The theory was, in fact, proposed as relevant to 
modelling quantum Hall states: although it was shown in \cite{NOI} that it fails 
to achieve one of its goal (the statistical transmutation of bosons on a line), 
the system presents some very interesting features. First of all, it can be 
exactly reduced, solving for $A_{\mu}$ and $B$, to a self-interacting bosonic
theory for which a local Lagrangian formulation is possible. Remarkably, the 
system possesses the non-relativistic scale invariance and for one
choice of the 
parameters we recover an integrable equation (DNLS): for a generic choice 
we have a scale-invariant perturbation  of the integrable model. 
The one-soliton structure of the theory has been examined, and it exhibits an 
interesting ``chiral''behaviour: one-soliton solutions exist only for a 
fixed sign of the total momentum and they present a very peculiar particle-like 
energy-momentum relation inherited from the scale invariance. Solutions 
with non-trivial boundary conditions at infinity (dark solitons) were
also found,
 existing for the opposite sign of momentum and with finite energy. In this 
case some care was needed in defining the conserved quantities: in particular 
we have proposed a definition of the total momentum consistent with the 
Hamiltonian structure of the theory and leading to a finite result. The scale 
invariance is broken by our boundary conditions, which entails a complicate 
dispersion relation. Finally we have studied the quantum dynamics, trying to 
discuss the relation between the classical one-soliton solutions and the 
quantum bound-states. The lack of integrability has not allowed a complete 
solution, but some result were obtained. The two-body problem was 
solved and the quantum bound states reproduce the chiral behaviour of the 
classical solutions. Moreover the perturbative computation of the energy for 
the N-body bound-state was found to be in agreement with the Bohr-Sommerfield 
quantization of the soliton in the weak-coupling limit. On the other hands 
scale invariance was broken at quantum level by a trace 
anomaly: the fixed point of the renormalization group coincides with the 
integrable system, while, in the general case, we have deduced the functional 
form of the bound-states energy by combining the (classical) one-soliton solution 
with the informations coming from the (quantum) trace anomaly. There remains to 
prove or, at least, to check our proposal by explicit computations: 
for example to calculate the functional determinant representing, in 
the path-integral 
framework, the non-trivial quantum corrections or to solve explicitly the 
three-body Schr\'odinger equation. 
This subject together with the non-abelian extension of the model are 
currently under investigations.

\section*{Acknowledgements}

It is a pleasure to acknowledge several suggestions and critical
readings from Professor Stanley Deser and Professor Roman Jackiw. A
warm thankyou goes to Dr. Ugo Aglietti who partecipated at the early
stages of this work.

\end{document}